\documentclass{WileyMSP-template}
\usepackage{textcomp}   
  
\usepackage{gensymb}    
\usepackage{parskip}
\usepackage{hyperref}
\usepackage{amsmath}
\usepackage[final]{changes}
\usepackage{multibib}
\newcites{SI}{References (Supporting Information)}

\begin{document}

\sloppy
\pagestyle{fancy}

\rhead{\includegraphics[width=2.5cm]{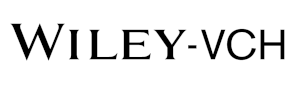}}

\title{Inverse-Designed Superchiral Hot Spot in Dielectric Meta-Cavity \replaced{Toward}{for} Ultra-Compact Enantioselective Detection}
\maketitle

\author{Anastasia Romashkina$^{\#}$*}
\author{Omer Yesilurt$^{\#}$}
\author{Vahagn Mkhitaryan}
\author{Owen Matthiessen}
\author{Min Jiang}
\author{Evgeny Lyubin}
\author{Bayarjargal N. Tugchin}
\author{Isabelle Staude}
\author{Jer-Shing Huang}
\author{Thomas Pertsch}
\author{Alexander V. Kildishev*}

$^{\#}$Contributed equally

\begin{affiliations}
Dr. O. Yesilurt, Dr. V. K. Mkhitaryan, O. M. Matthiessen, Prof. A. V. Kildishev*\\
Elmore Family School of Electrical and Computer Engineering, Birck Nanotechnology Center, and Purdue Quantum Science and
Engineering Institute, Purdue University, West Lafayette, IN 47907, USA \\ 
Email Address: kildishev@purdue.edu \\

A. M. Romashkina*, Dr. E. V. Lyubin, Dr. B. N. Tugchin, Prof. I. Staude, Prof. T. Pertsch\\
Institute of Applied Physics, Abbe Center of Photonics, Friedrich Schiller University Jena, Albert Einstein Str. 6, 07745 Jena, Germany \\
Email Address: anastasia.romashkina@uni-jena.de
\\
M. Jiang, Prof. J.-S. Huang\\
Research Department of Nanooptics, Leibniz Institute of Photonic Technology, Albert Einstein Str. 9, 07745 Jena, Germany
\\
Prof. I. Staude \\
Institute of Solid State Physics (IFK), Friedrich Schiller University Jena, Helmholtzweg 3, 07743 Jena, Germany
\end{affiliations}

\keywords{optical chirality, superchiral cavity, dielectric metasurface, inverse design, topology optimization, ML-assisted design}

\begin{abstract}
Chiral nanophotonic structures have garnered considerable interest in recent years due to their potential to enhance the efficacy of chirality-sensitive biomolecular detection.
Designing metaplatforms to enhance chiroptical signals under linearly polarized excitation is particularly appealing due to the minimal chiral background and the ease of controlling excitation polarization.
Here, a novel two-step inverse design scheme for dielectric lossless metasurfaces with superchiral hot spots is proposed. 
The method extends the local density of field enhancements for non-chiral fields into the chiral regime and significantly surpasses previous enhancements in superchiral field generation. 
It has been demonstrated that by leveraging the excitation of high quality factor modes with small mode volumes, it is theoretically possible to convert linearly polarized plane waves into a superchiral hot spot with record-high enhancement in the near-field optical chirality up to 10$^4$.
A prototype is successfully implemented using advanced nanofabrication technologies. The optical characterization of the prototype demonstrates a $10^2$-fold enhancement in optical chirality.
The findings of this study unveil novel prospects for chiral spectroscopy with ultra-compact devices, underscoring the role of machine learning and physics-based inverse design in the development of cutting-edge, functional photonic structures.

\end{abstract}

\section{Introduction}
Chirality, or the lack of mirror symmetry, is a widespread geometric property that appears commonly in the natural world and reflects nature's inherent bias towards asymmetry~\cite{masonOrigin1986}. 
In particular, it occurs at the molecular level, when the molecules become fundamentally distinct and non-superimposable with their mirror image~\cite{cahnSpecification1966}. 
Many biological molecules exhibit equivalent chemical structures and physical properties but significantly distinct biological effects between different enantiomers~\cite{nguyenChiral2006}.
This necessitates the development of precise techniques that can carefully distinguish molecular enantiomers with high sensitivity for use in the pharmaceutical industry, healthcare, and other related fields.

In accordance with their physical asymmetry, these chiral biomolecules exhibit divergent electromagnetic responses to circularly polarized optical excitation of opposite rotation, allowing themselves to be detected optically~\cite{fasmanCircular1996a}.
Circular dichroism (CD) spectroscopy is an analytical technique that utilizes the difference in absorption between left- and right- circularly polarized light~\cite{beychokCircular1966}.
However, this absorption difference is extremely small for many naturally occurring molecules, and therefore optical detection remains a significant challenge.

Here, nanophotonic platforms offer a solution to enhance the chiroptical response of enantiomers by generating electromagnetic fields with low mode volume and a high density of states.
Mostly, the efforts are concentrated around metasurfaces as artificially engineered planar materials with unique functionalities~\cite{kuznetsovRoadmap2024, minovichFunctional2015}.
In this context, a promising strategy relies on enhancing the local optical chirality $C$, which is defined in time-averaged form for a monochromatic wave as~\cite{yooNanophotonic2025}
\begin{equation}
    C = \frac{\omega\varepsilon_0}{2} \mathrm{Im}(\mathbf{\tilde{E}}
    \cdot\mathbf{\tilde{B}^*}),
    \label{eq:chirality}
\end{equation}
where $\omega$ is the angular optical frequency, $\varepsilon_0$ is the vacuum permittivity, and $\mathbf{\tilde{E}}$ and $\mathbf{\tilde{B}}$ are the time-averaged electric and magnetic field vectors, respectively. 
Then, the differential optical absorption ($\Delta A$) spectroscopy can be performed, as it was shown that the molecular absorption rate difference for left- ($A^-$) and right- ($A^+$) circularly polarized light for a chiral molecule is proportional to the local optical chirality density $C$, via
\begin{equation}
    \Delta A = A^+ - A^- = 2 G''\omega \mathrm{Im}(\mathbf{\tilde{E}}
    \cdot\mathbf{\tilde{B}^*}),
\end{equation}
where $G''$ is the isotropic mixed electric-magnetic polarizability \cite{tangOptical2010}.
This provides a means to enhance the naturally weak chiroptical response by increasing optical chirality. Furthermore,
\replaced{optical trapping can add extra functionality to a chiral sensor~\cite{lin2021plasmonic}}{the high density of electromagnetic states enables the optical trapping and manipulation of biomolecules~\cite{huangOrigin2015}}.

For these purposes, chiral plasmonic structures were extensively investigated due to their ability to deliver a strong concentration of electromagnetic energy to subwavelength volumes~\cite{hentschelChiral2017}. 
Numerous studies have proposed chiral-sensing platforms based on plasmonic metasurfaces~\cite{helgertChiral2011, shaltoutOptically2014, linSlantgap2014, yinActive2015, fasoldDisorderEnabled2018, tabouillotNearField2022, biswasTunable2024}.
However, the strong absorption of light by metals makes the plasmonic platform less suitable for application in photonic devices and limits the potential enhancement factors to 10-100~\cite{westSearching2010, schullerPlasmonics2010, choudhuryMaterial2018}. 

High refractive index dielectric nanophotonic devices, in contrast, can exhibit very low optical losses at optical frequencies~\cite{deckerResonant2016}. 
Various shapes of mainly silicon resonators were used to construct chiral metasurfaces~\cite{maAlldielectric2018, solomonNanophotonic2020, wangSimultaneous2023}.
High-refractive-index dielectric resonators support both electric and magnetic multipolar Mie resonances, enabling single-sign enhancements of $C$ ~\replaced{\cite{solomonNanophotonic2020, garcia2019enhanced}}{\cite{solomonNanophotonic2020}}. 
However, there is always a compromise between strong enhancement with higher refractive index and optical absorption.
High-quality-factor (high-Q) resonant enhancement of chiroptical response is usually achieved by careful engineering of surface lattice resonances~\cite{luo_high-q_2023, toftul_chiral_2024} or bound state in the continuum (BIC) resonances~\cite{hao_enhanced_2025, wei_quasi-bic-enhanced_2025, guan_ultrasensitive_2025}.

A considerable number of reported metasurface sensing platforms depend on the achiral meta-atoms and circularly polarized light, which generally necessitates the incorporation of additional optical elements~\replaced{\cite{garcia2019enhanced, deng2024advances}}{\cite{deng2024advances}}. 
An alternative approach involves the design of a structure capable of directly converting linearly polarized incident light into a hotspot with enhanced optical chirality of one handedness, which would significantly facilitate the design of a compact device.
While the designs utilizing the angled incident light are intensively studied \cite{plum2009metamaterials, lin2021plasmonic, singh2023dielectric}, the well-known efficient intuitive geometries creating superchirality under normal incidence remain limited \cite{baeInverse2025}.

In recent years, photonic inverse design has emerged as a promising strategy and tool for achieving substantial enhancement of various functionalities and highly localized fields~\cite {moleskyInverse2018, liEmpowering2022}. In contrast to the forward design, which uses parameter-based optimization approaches, it enables the discovery of non-intuitive geometries by sampling from a much higher-dimensional design space~\cite{kudyshevMachine2021, kuhnInverse2022, gladyshevInverse2023}. A representative example of such a high-dimensional inverse design framework is topology optimization (TO)~\cite{bendsøeTopology2004, jensenTopology2011, liTopology2019}. TO belongs to the class of high-dimensional gradient-descent techniques and is used to find the ultimate volume-discretized material distribution within a specified design domain, thereby achieving optimal performance under predefined constraints~\cite{christiansenInverse2021, baeInverse2025}.  Despite its large design parameter space, defined by numerous discrete material elements, TO achieves outstanding numerical efficiency, ensured by the adjoint method for computing high-dimensional derivatives. Since adjoint optimization is generally a local method, restarting with random seeds is often used to avoid becoming trapped in a local minimum. To obtain a global solution, evolutionary and deep learning approaches have recently been employed to guide TO engines. 
Deep learning is rapidly advancing in the field of photonic design~\cite{maDeep2021}. Neural networks (NNs) are a popular tool to speed up the computationally expensive forward simulations \cite{tahersima_deep_2019, ma_pic2o-sim_2025,kang_adjoint_2025, lynch_physics-guided_2025} or find nontrivial design solutions \cite{jiang_free-form_2019, jiang_simulator-based_2020, dai_shaping_2025}. Using a combination of these methods, highly efficient structure geometries were found, e.g. for three-dimensional photonic crystals~\cite{menRobust2014}, electromagnetic field localization~\cite{vester-petersenTopology2017}, optical tweezers~\cite{nelsonInverse2024}, thermal emitters~\cite{kudyshevMachinelearningassisted2020}, nanolenses~\cite{augensteinInverse2020}, photonic couplers~\cite{yesilyurtEfficient2021}, metasurfaces~\cite{suTopology2024}, and meta-cavities~\cite{matthiessenTopologyOptimized2025}.
Recent work has demonstrated a dielectric optical cavity with mode volumes close to the fundamental limit~\cite{albrechtsenNanometerscale2022}.

Machine learning and deep learning methods are also becoming established tools to optimize the performance of chirality-targeting nanostructures~\cite{liaoDeeplearningbased2022, augensteinNeural2023, romashkinaCollective2023a, hanNeuralNetworkEnabled2023, qiuChiral2023}.
Some studies have shown that inverse design can enhance the quality factor of chiroptical metamaterials~\cite{baeInverse2025}. 
In this direction, TO was applied to inverse design a dielectric metasurface with strong helical dichroism \cite {panInverse2025}. 
A volumetric metamaterial design was proposed to achieve a superchiral nanoscale hotspot under circular light excitation~\cite{taguchiNanoscale2025}.

In this work, we apply TO to inverse design a highly efficient superchiral device based on a dielectric metasurface. 
In comparison to volumetric structures and metamaterials, which can show intrinsically high chiroptical response~\cite{menzelAsymmetric2010}, a planar metasurface platform allows for much simpler and scalable fabrication~\cite{kildishevPlanar2013, minovichFunctional2015}, while still allowing for chiroptical response because of air-substrate asymmetry~\cite{zhaoManipulating2011, khanikaevExperimental2016}.
\added{Here we focus on the enhancement of optical chirality density for linearly polarized illumination.}
By confining electromagnetic energy in both frequency and space, a 4 orders-of-magnitude enhancement of optical chirality is achieved in the designed structure. 
This is followed by fabrication and experimental demonstration of the structure performance, which, to the best of our knowledge, is the first real-scale experimental realization of an inverse-designed chiral structure generating superchiral fields~\cite{baeInverse2025}.
This work paves the way for single-molecular chiral sensing with compact, fabricable devices.

\section{Results and discussion}
\subsection{Design}
The present study explores the two-step inverse design of a superchiral dielectric metasurface. 
The designed metasurface is a biperiodic array of identical unit cells. An inverse design method using TO is applied to the unit cell to achieve a non-intuitive geometric pattern that produces a superchiral hotspot at a target wavelength of 1310~nm. 

\begin{figure}
  \includegraphics[width=0.96\linewidth]{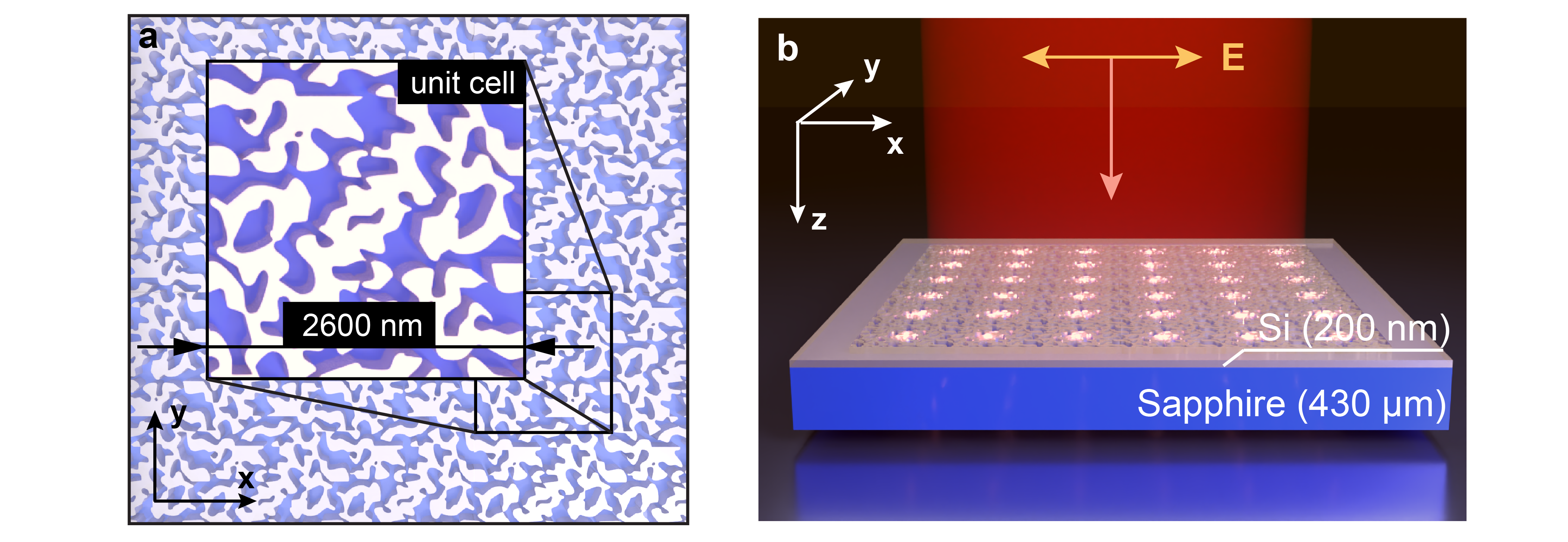}
  \caption{\textbf{Superchiral cavity design.} (a) Top view of the designed metasurface, with an inset showing the 2.6-$\micro$m-wide meta-unit cell.
(b) Schematic side view of the designed metasurface representing a 200-nm-thick patterned silicon film on a sapphire substrate. The structure is designed to be illuminated under normal incidence with a plane wave propagating along the $z$-axis with linear polarization along the cavity (along $x$-axis). Horizontal (H) and vertical (V) polarizations correspond to the electric field vector orientation along $x$- and $y$- axes, respectively}
  \label{fig:design}
\end{figure}

\textbf{Figure~\ref{fig:design}a,b} show the top and side views, respectively, of the proposed metasurface design. The metasurface consists of a structured 200-nm thick silicon film on a sapphire substrate. 
The silicon edges are modeled as vertical walls. The pattern has a period of 2.6~$\micro$m, which is almost twice the operation wavelength of the metasurface of 1310~nm. 
The metasurface is illuminated by a plane-wave source with linear polarization at surface-normal incidence. 

The metasurface is optimized to generate very tight, collocated electric and magnetic hotspots resulting in a superchiral "near field".
The figure of merit (FoM) of the optimization procedure is the absolute value of the optical chirality ($C$ in Equation~\ref{eq:chirality}) at the center of the unit cell on the surface of the structure, when the metasurface is illuminated with \replaced{}{a} linearly polarized light \cite{yesilurtPHYSICS2024}.

\begin{figure}
  \includegraphics[width=0.89\linewidth]{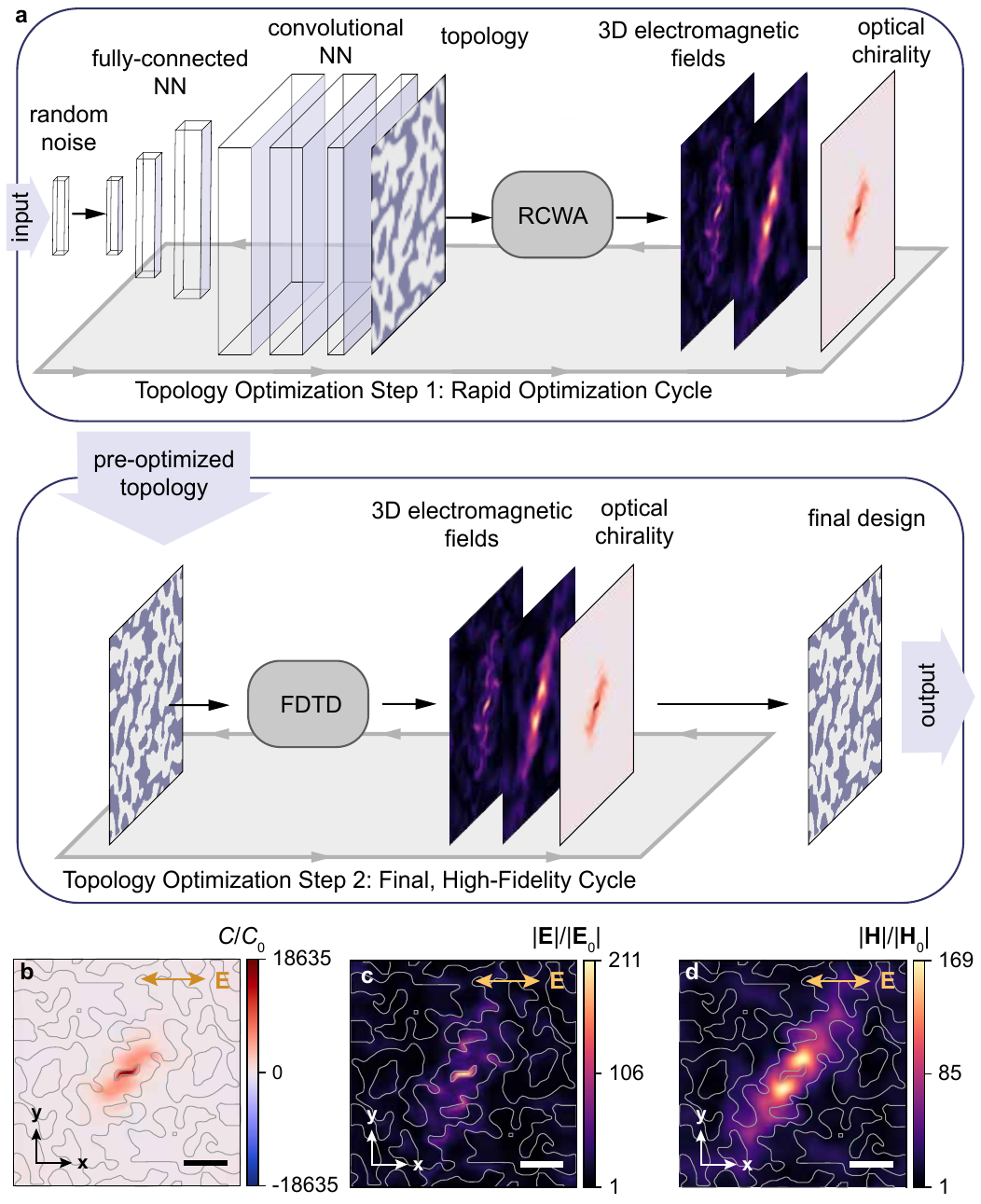}
  \caption{\textbf{2-step inverse design method.} In the first step, the random noise vector is propagated through fully connected neural network (NN) layers, and its dimensions are expanded with convolutional layers. The NN output is a binary, pixelated image representing the Si-pattern shape in the unit cell of the metasurface. The optical response of the metasurface is calculated with RCWA. Using the stochastic gradient optimizer with fully differentiable RCWA, the NN weights are optimized to achieve maximum chirality~$C$ enhancement at the center of the unit cell. In the second step, the pre-optimized topology is further fine-tuned using an adjoint-variable method with the direct FDTD solver.
  (b-d) Simulated pseudo-color maps over the top surface of the unit cell, showing the normalized near-field optical chirality density $C$ (b), and the normalized magnitudes of the $\mathbf{E}$ and $\mathbf{H}$ near-fields  (c), (d), respectively, under monochromatic plane wave excitation at a wavelength of 1310 nm. All maps are normalized to those for circularly polarized light propagating in free space. The gray lines outline the shape of the metasurface pattern. Scale bar is 500~nm.}
  \label{fig:NN}
\end{figure}

\textbf{Figure~\ref{fig:NN}a} illustrates the schematics of the inverse design procedure.
The optimization process was divided into two steps. 
The first (preliminary) step employed a fully differentiable Rigorous Coupled Wave Analysis (RCWA) method, and optimization was performed via stochastic gradient descent to avoid trapping in the local minima. The primary aim of this rapid preliminary step was to optimize NN weights rather than the topology itself, thereby enabling higher-dimensional search in the design space. \added{A 180-degree rotation symmetry was imposed onto the topology to ensure smooth convergence and easier manufacturing.}
At the second step, the optimization using the Finite-Difference Time-Domain (FDTD) numerical method was employed to mitigate the known limitations of the RCWA method in accurately reconstructing high-contrast refractive-index jumps.
There, an adjoint-optimization method was used to fine-tune the pre-optimized geometry to achieve more accurate performance predictions. 
Within that optimization loop, additional filtering was applied to ensure robustness against fabrication imperfections~\cite{häggMinimum2018}.

\added{The fully differentiable RCWA implementation can suffer from Gibbs-type artifacts arising from truncated Fourier representations at high-contrast material boundaries, which can reduce the accuracy of strongly localized near-field quantities. However, directly replacing the first step by FDTD would not provide the same end-to-end rapid and global differentiable NN-driven exploration in our current framework and would require substantially more computationally expensive adjoint-FDTD simulations from many initial topologies.}
Further details on the optimization algorithm are given in the \textbf{Methods} section.

\added{The optimized figure of merit was $\mathrm{FoM}=|C_H(x_0,y_0,z_0;\lambda_0=\mathrm{1310~nm})|$, where H denotes the incident linear polarization along $x$-axis.}
The normalized chirality $C/C_0$, where $C_0$ is the chirality of a circularly polarized plane wave in free space of the same power as the one illuminating the structure, is enhanced up to $1.8\cdot10^4$ inside the cavity on the surface of the structure (Figure~\ref{fig:NN}b). 
This result is supported by the simultaneous enhancement of the electric field $|\mathbf{E}|/|\mathbf{E_0}|$ (Figure~\ref{fig:NN}c) and the magnetic field $|\mathbf{H}|/|\mathbf{H_0}|$ (Figure~\ref{fig:NN}d) near the center of the cavity. 
Here, $\mathbf{E_0}$ and $\mathbf{H_0}$ are the electric and magnetic field intensities for the circularly polarized plane wave propagation in free space.

We present spectrally resolved resonance characterization based on FDTD simulations in the Supplementary Information. This includes \replaced{zeroth}{first} diffraction order transmission spectra for vertical, horizontal, right- and left-circular incident polarizations, as well as CD in the transmission spectrum. 
It was previously theoretically shown that the third Stokes parameter can be used as a far-field-observed characteristic of the near-field optical chirality \cite{poulikakosOptical2016}. Therefore, we extend our characterization by adding the FDTD-simulated spectra of the third Stokes parameter for the incident polarization mentioned above.
\subsection{Fabrication and characterization}
We fabricate the metasurface using top-down fabrication technology with electron beam lithography. 
Details of the fabrication procedure are provided in the Methods section. The transferred pattern has a period of $2.6~\micro$m in the $x$ and $y$ directions. The thickness of the silicon film in the investigated sample is 210~nm. 

To characterize the resonant chirooptical response of the metasurface spectrally, we perform transmittance spectroscopy. A detailed description of the experimental setup is given in the Methods section.

\begin{figure}
\includegraphics[width=0.9\linewidth]{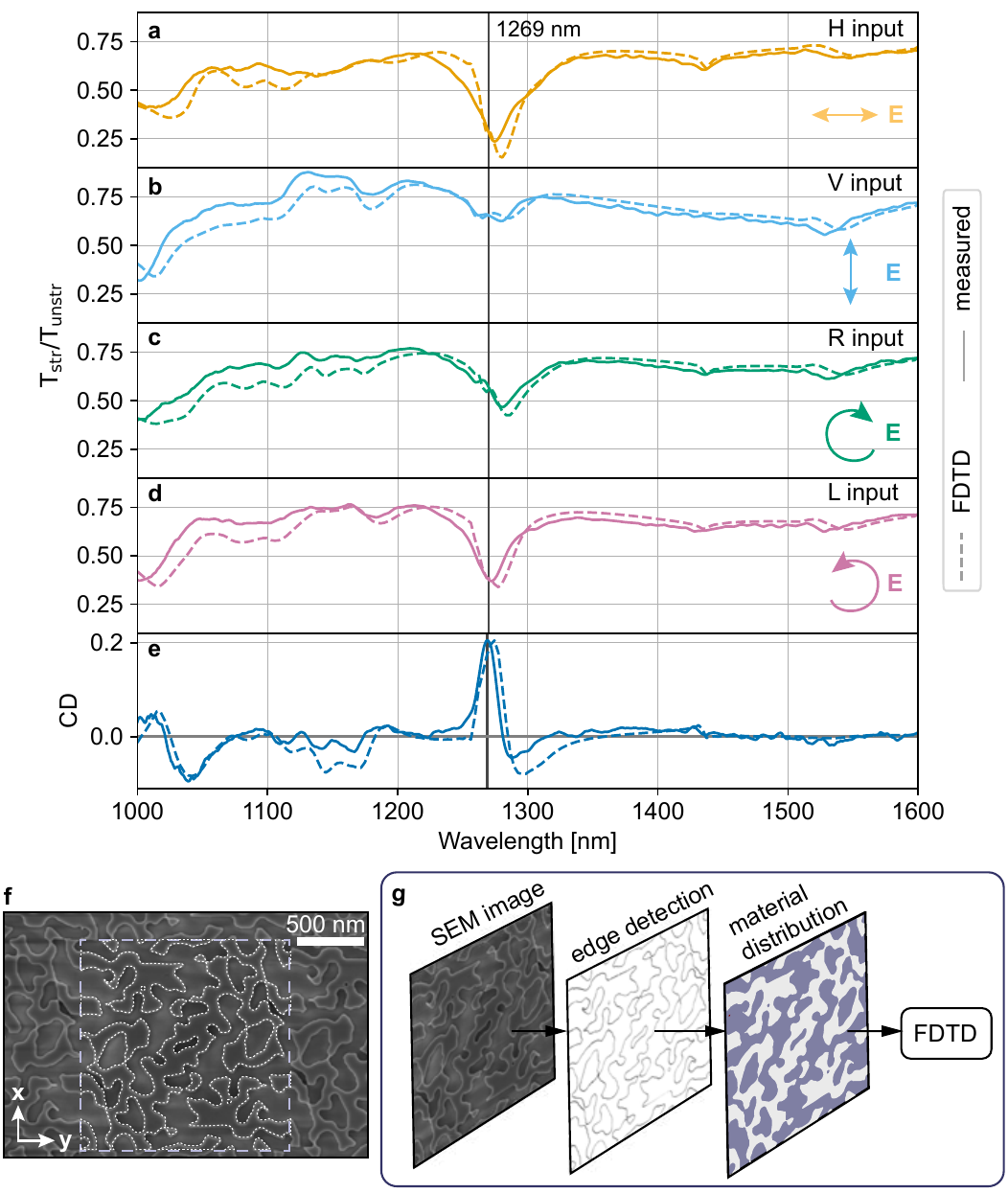}
  \caption{\textbf{Fabricated sample and experimental results.} (a-e) The experimental (solid line) and FDTD-simulated (dashed line) spectra of the total transmission of the metasurface under a plane wave illumination at normal incidence with H- (a), V- (b), R- (c), and L- (d) polarization. The spectra are normalized to the transmission spectra of an unstructured Si film on the same substrate. (e) The experimental (solid line) and FDTD-simulated (dashed line) CD spectra in transmission, calculated from Equation (\ref{eq:cd}). Gray vertical line marks 1270 nm wavelength.  (f) SEM image of the fabricated metasurface. Dashed lines outline the unit cell of the periodic metasurface. Dotted lines highlight the extracted edges of the patterned silicon film. (g) Schematic representation of the method used to build the FDTD model of the fabricated structure.}
  \label{fig:spectra}
\end{figure}

\textbf{Figure~\ref{fig:spectra}a,b,c,d} shows the experimental (solid lines) transmission spectra of the sample. The total transmitted light intensity is measured for horizontal (H), vertical (V), right-(R), and left-(L) circular polarizations and normalized to the respective intensity of the transmitted light through an unstructured film of the same thickness on the same substrate. 

In contrast to the transmission spectra for V-polarization, the H-polarization transmission spectrum shows a prominent "dip" centered at 1271~nm.
From the difference between the total transmission of the right- and left- circular\added{ly} polarized light, the CD in transmission is calculated as
\begin{equation}
    \mathrm{CD} = \frac{T_\mathrm{R}-T_\mathrm{L}}{T_\mathrm{R}+T_\mathrm{L}},
    \label{eq:cd}
\end{equation}
where $T_\mathrm{R}$ and $T_\mathrm{L}$ are the transmission coefficients of R- and L- circularly polarized light, respectively.
\textbf{Figure~\ref{fig:spectra}e} shows the CD spectrum calculated with Equation (\ref{eq:cd}).
A strong resonant enhancement of the CD up to the value of 20\% is observed at 1269~nm. 

The direct characterization of the hotspot and its field enhancement, as, e.g. by scanning near-field optical microscopy (SNOM), remains challenging
due to the sensitivity of the high-Q resonance to the angle and the numerical aperture (NA) of illumination, as well as perturbation by the SNOM tips.
Therefore, we utilize a non-invasive method to evaluate the performance of freeform-optimized metasurfaces with nontrivial geometries. This method combines far-field spectroscopy with fabrication-aware numerical modeling.

\textbf{Figure~\ref{fig:spectra}f} shows the top-view Scanning Electron Microscopy (SEM) image of the fabricated structure.
\textbf{Figure~\ref{fig:spectra}g} shows the methodology for modeling the fabricated structure. 
We process the SEM image to extract the edges of the patterned silicon film. Afterwards, we transform these edges into a binary image, which we then insert into the FDTD model (Ansys Lumerical). 
Due to the limited resolution of the SEM imaging, electron scattering artifacts, and the natural oxidation of Si by up to 5-10 nm, the dimensions of silicon edges can vary slightly. Therefore, we performed simulations with controlled lateral offsets of the silicon edge positions to determine the exact dimensions.
The model is designed to closely simulate the experimental observables of far-field optical characterization measurements.
The detailed model parameters are provided in the \textbf{Methods} section.

Figure~\ref{fig:spectra}a,b,c,d shows the numerically calculated (dashed line) and measured (solid line) far-field transmission spectra of the sample. The total transmitted light intensity is obtained for H-, V-, R-, and L-polarizations and normalized to the intensity transmitted through the unstructured film of the same thickness on the same substrate. 
The experimental and numerically modeled spectra show excellent agreement, and the numerically modeled spectra exhibit a slight redshift of 5~nm, which might be due to an oxidation shell on the Si that is not visible on the SEM images. 
As shown in Figure~\ref{fig:spectra}e, the simulation produced a maximum CD value of 20\%, which is the same as the measured value. 
\textbf{Figure~\ref{fig:ff_nf_sp}a} shows measured (solid line) and simulated (dashed line) CD in the narrow spectral region from 1240~nm to 1310~nm. 
The central wavelength of the CD resonance is at 1269~nm for measured and 1274~nm for the numerically calculated spectra.
The full width at half maximum (FWHM) values are also very similar, with 15~nm for the measured case and 18~nm for the simulation. 
The quality factor of the resonance is calculated from the FWHM values, which are 70 and 83 for the simulated and measured cases, respectively.

\begin{figure}
\includegraphics[width=0.9\linewidth]{figures/fig_cd_nf.pdf}
  \caption{\textbf{Resonance spectral characterization.} (a) The experimental (solid line) and FDTD-simulated (dashed line) CD spectra in transmission, calculated with Equation~\ref{eq:cd}; (b) Spectra of maximum local chirality value (pink line) and average local chirality (light blue line) across the unit cell on the surface of the structure. The values are normalized to the local chirality value for a circularly polarized plane wave in free space. 
  }
  \label{fig:ff_nf_sp}
\end{figure}

\begin{figure}
  \includegraphics[width=0.96\linewidth]{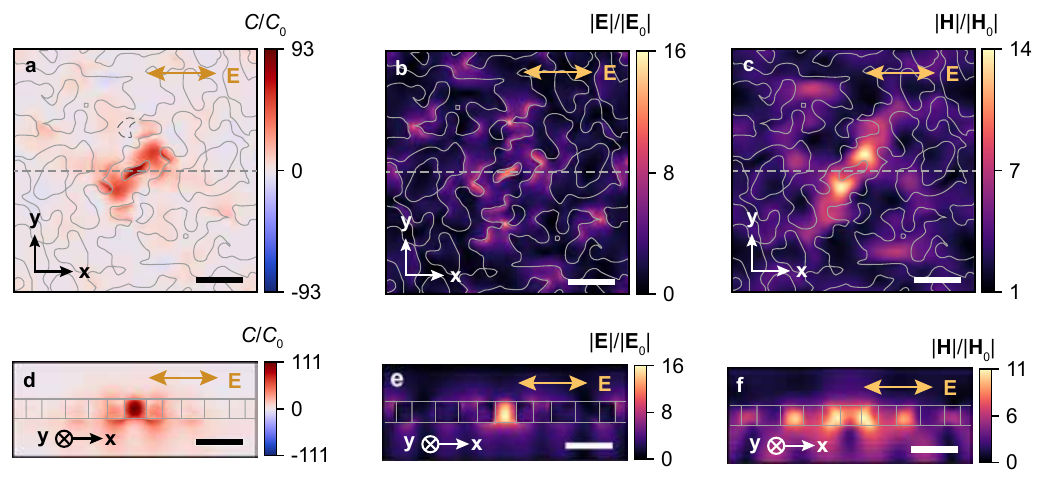}
  \caption{\textbf{FDTD-simulated near fields of the fabricated metasurface.} (a-c) Simulated pseudo-color maps over the top surface of the unit cell, showing the normalized near-field optical chirality $C$ (a) and the normalized magnitudes of the E and H near fields  (b and c). The gray dashed line shows the location of the $xz$ plane in (d-f). (d-f) Simulated pseudo-color maps in the $xz$ plane crossing the center of the cavity, showing the normalized near-field optical chirality $C$ (d) and the normalized magnitudes of the E and H near fields (e and f), under monochromatic plane wave excitation at a wavelength of 1267~nm. All maps are normalized to those for circularly polarized light propagating in free space. The gray lines outline the shape of the pattern. The scale bar is 500~nm.}
  \label{fig:nf}
\end{figure}

The near-field characteristics of the fabricated structure are evaluated numerically using the FDTD method. 
The parameters of the FDTD model are the same as those used to simulate the far-field spectra, except that a narrowband source is employed instead of a broadband source to enable comparison with the design, which was optimized for monochromatic excitation. 

We sweep the excitation center wavelength in steps of 1~nm to find the maximum chirality enhancement. 
\textbf{Figure~\ref{fig:ff_nf_sp}b} shows the maximum local optical chirality $C$ on the surface of the structure (pink line) and the average optical chirality $C$ on the surface of the structure (light blue line) plotted against the wavelength of the narrowband illumination.
All values are normalized to the respective values extracted from the reference simulation with a circularly polarized plane wave of the same power propagating in free space.
The maximum values of both occur at an illumination wavelength of 1267~nm.
The FDTD-simulated CD exhibits a maximum at 1274~nm, which is redshifted by 7~nm compared to the spectral position of the near-field maximum enhancement~\cite{menzelSpectral2014}.
Therefore, the resonant optical chirality enhancement in the fabricated structure is predicted to occur at a wavelength \replaced{near 1262}{of 1260}~nm\added{, blueshifted by 7~nm relative to the experimental CD maximum}.

\textbf{Figure~\ref{fig:nf}} shows numerically calculated pseudo-color maps of the optical chirality $C$, and electric and magnetic fields.
A strong enhancement of optical chirality is clearly visible in the gap at the center of the unit cell (Figure~\ref{fig:nf}a).
The simultaneous enhancement of electric and magnetic fields in the vicinity of the gap supports this enhancement.
The enhancement ratio with respect to circularly polarized light is approximately 100.
\added{Section S7 of the Supporting Information discusses how small variations in the design geometry affect the resulting enhancement values, showing that fabrication-induced deviations may lead to a significant reduction in the achievable enhancement.}
\replaced{Furthermore, w}{W}e note that the actual $C$ enhancement factor may be limited by imperfect periodicity, which is not accounted for in the ideally periodic FDTD simulation model.

\added{Here we note that in a practical compact sensing scheme with a linear excitation, the opposite-handed channel would be provided by a mirror-enantiomeric meta-cavity patch on the same chip illuminated with the same linear polarization and functionalized with the same molecular concentration.}

To further characterize the polarization state of the light after interaction with the metasurface, we perform polarization-resolved transmission spectroscopy. 
The Stokes vector components under a plane wave illumination at a normal incidence for 4 incident polarization states were recorded, and the results are presented in the Supplementary Information, along with the corresponding components obtained from FDTD simulations.
\section{Conclusion}
We experimentally demonstrated a superchiral silicon-on-sapphire metasurface device prototype with a local chirality enhancement of $\approx$100 times that of circularly polarized light in vacuum. 
A two-step inverse design framework, combining rapid RCWA-based neural-network exploration with an adjoint FDTD refinement step, was employed to inverse-design the structure. 
While the method predicts that such superchiral metasurface-based devices can potentially achieve chirality enhancement on the order of $10^4$, our study demonstrates that inverse-designed dielectric architectures can already achieve levels of experimental superchiral field enhancement previously accessible only with lossy plasmonic systems. 
Unlike metallic platforms, where absorption limits both the achievable quality factor and the magnitude of chirality, the silicon-on-sapphire cavity supports low-loss, high-Q resonances with subwavelength mode confinement that is constrained by fabrication limitations rather than intrinsic material properties.
The study highlights a deep-learning-assisted two-stage inverse design strategy as a promising utility for developing miniature, real-world chiral sensing devices. 
\added{Further design development of the chiral sensor would benefit from fully switchable chirality enhancement behavior of one geometry for realizing ultra-compact chiral sensing. In the real device, further consideration should be taken about the average optical chirality enhancement and background subtraction.}
\section{Methods}

\threesubsection{Two-step topology optimization}
In the first step, the topology optimization was driven by an NN. The NN progressively expanded a noise vector into a complex topology of the correct dimensions. \added{More specifically, the NN was a generative decoder comprising 13 trainable layers: 10 fully connected layers and three convolutional layers, in addition to two LayerNorm and two BatchNorm2d modules. The noise vector was initialized once and remained fixed, while the trainable weights and biases of the decoder were optimized. The detailed structure of the NN is presented in Section S8 of Supporting Information.} The electromagnetic response of the proposed design was calculated using the RCWA method. \added{RCWA simulations were performed using torcwa, a PyTorch-based differentiable RCWA implementation \cite{torcwa}.} After the loss was obtained, the gradients were back-propagated through the fully differentiable RCWA into the NN to update the parameters. \added{These parameters were updated using the Adam optimizer. Thus, the topology was modified indirectly through changes in the output of the decoder, rather than being optimized directly as an independent set of design variables.} This process was iterated until the FoM converged or the user terminated it. 
In the second step, the FDTD solver was used to optimize the topology obtained in the preliminary step. \added{FDTD simulations were performed using Ansys Lumerical FDTD through the lumapi Python interface \cite{Lumerical}.} The TO gradients were derived using the adjoint variable method. At each iteration, the calculated fields within the three-dimensional cavity volume were first reduced to a two-dimensional matrix. That was achieved by averaging the fields along the propagation direction in the FDTD grid. The averaging step ensured uniform material composition along that direction. The averaged fields were then used to compute gradients, which were scaled and added to the index-density matrix to update the cavity topology. To ensure compatibility with advanced fabrication methods, such as electron-beam lithography, a circular spatial blurring filter with a radius of 40 nm was periodically applied to eliminate sub-threshold features in the material distribution. Though that process initially reduced the FoM, it was repeated until a design emerged that was minimally affected by the filtering. To achieve a fully binarized refractive index distribution, a Heaviside filter was employed. Because the second stage started from a pre-optimized topology, 200 iterations were sufficient.

\threesubsection{Fabrication procedure}
Top-down nanofabrication methods were performed on a silicon-on-sapphire wafer via electron beam lithography (EBL). 
A 200 nm-thick epitaxial silicon-on-sapphire (SoS) wafer was diced into 6.5~by~6.5 mm pieces. 
The pieces were then cleaned by ultrasonication in acetone, isopropanol, and deionized water, and dried with dry N$_2$. 
The sample surface was dehydrated by heating under nitrogen and primed with hexamethyldisilazane vapor in a Yield Engineering Services YES~LP-III vacuum oven.
The sample was spin-coated with a 50 nm thick film of EBL resist (ZEP~520A diluted 1:2 in anisole) at 4000~rpm for 45~s, and the coated sample was baked on a hotplate at 180$^o$C for 3~min. 
100 by 100~$\micro$m$^2$ meta-cavity arrays are written in the electron beam resist using an Elionix ELS-BODEN 150 EBL system with a 150~keV electron beam at a current of 200~pA and base exposure dose of 325 $\micro$C/cm$^2$. 
To address electron scattering during exposure, which can distort small features, proximity effect correction (PEC) algorithms were employed. 
These algorithms adjusted the dose and geometry of the patterns to compensate for electron-beam spreading and improve the fidelity of the final design. 
The resist development process was fine-tuned to minimize the impact of proximity exposure and enhance the resolution of the final features. 
Multiple copies of the array were fabricated with various combinations of unit-cell scaling (detuning) and PEC parameters. 
The sample was developed in chilled (approx. 3$^o$ C) o-xylene for 30~s, rinsed in isopropanol for 30~s, and dried with dry N$_2$.
The developed pattern is transferred to the underlying silicon by inductively coupled reactive ion etching in an SPTS Omega LPX Rapier reactive ion etcher. 
After etching, the remaining electron beam resist was removed by rinsing in acetone and isopropanol, followed by 4~min of O$_2$ plasma etching in a Matrix Integrated Systems 105 plasma asher.

\begin{figure}
  \includegraphics[width=\linewidth]{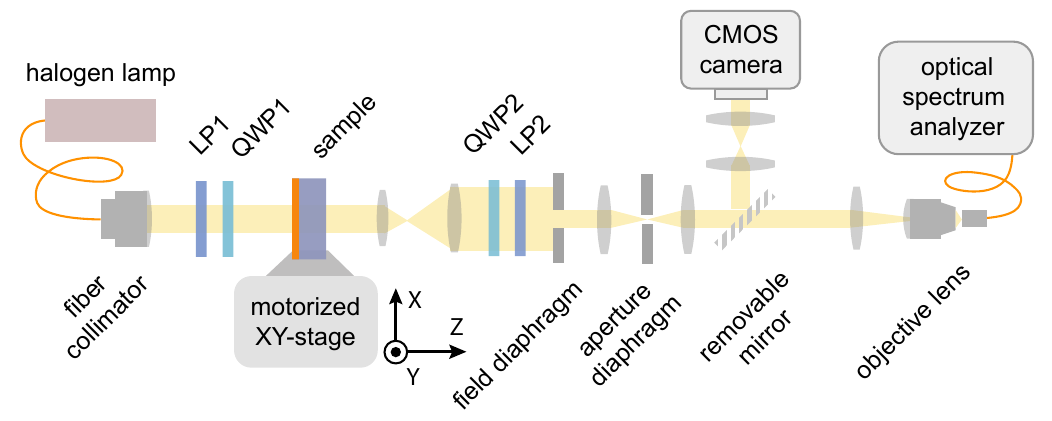}
  \caption{\textbf{Experimental setup.} The experimental setup for measuring the polarization-resolved transmission spectra of the sample.}
  \label{fig:setup}
\end{figure}

\threesubsection{Experimental optical characterization}
Figure \ref{fig:setup} shows the experimental setup used to measure the polarization-resolved transmission spectra of the sample.
The light from a halogen lamp (SLS301, Thorlabs) is collimated and transmitted through a combination of a linear polarizer (LP1) and an achromatic quarter-wave-plate (QWP1) to select the desired input polarization.
The sample is mounted on a motorized XY stage that allows for automatized collection of the signal from different areas of the sample (in particular, structured and unstructured silicon film). The sample is illuminated under normal incidence. 
The forward-scattered light is collected by a lens. 
A sequence of a quarter-wave-plate (QWP2) and a linear polarizer (LP2) enables the separate detection of H-, V-, L-, or R-polarized light. 
For the total transmission measurements, these two components are removed.
With the two-axis adjustable slit (field diaphragm) installed in the first image plane, we select the light transmitted only from the area of interest on the sample for the measurement. 
The aperture diaphragm limits the NA to near-normal incidence. 
The removable mirror allows for switching between two light paths: one towards the spectrometer (Optical Spectrum Analyzer AQ6370B, Yokogawa) and the other to a camera that is used to select the area of interest on the sample. 

\threesubsection{Numerical model of the fabricated sample}
We numerically modeled the fabricated sample using the FDTD method in Ansys Lumerical. \added{The model used the same FDTD simulation settings as the design-step FDTD optimization.}
The metasurface was modeled as an infinitely extended patterned silicon film by applying periodic boundary conditions along $x$- and $y$-axis to the boundaries of the 2,600-nm-square unit cell. The material data for silicon and sapphire were obtained using ellipsometry measurements. 
The thickness of the silicon film \replaced{was}{is} \added{set to}~210 nm. \added{Although the nominal Si layer thickness of the silicon-on-sapphire wafer was 200~nm, the thickness tolerance was $\pm$12\%, and the thickness measured during fabrication was approximately 210~nm. This value was therefore used in the post-fabrication FDTD model, where it also gave better agreement with the measured spectra than the nominal thickness.}
We used a normally incident plane-wave source, with a wave propagating along the $z$-axis. 
We simulated the spectral response with a broadband source with a bandwidth from 1000 to 1600~nm.
To simulate the near-fields, a narrowband source was used.
Along the $z$-axis, we applied stretched coordinate perfectly matched layer (PML) boundary conditions with a steep angle profile. The pattern geometry was imported using the image import utility. The maximum mesh step inside the Si film was set to be not more than 10 nm along $x$-, $y$-, $z$- directions.
2D frequency-domain field-profile monitors were used to record electric and magnetic field distributions. For obtaining the reference maps ($C_0$, $|\mathbf{E_0}|$ and $|\mathbf{H_0}|$), a simulation with a circularly polarized reference field propagating in vacuum is performed. The reference field was generated with a circularly polarized source of the same power.
To match the experimental geometry of far-field spectroscopy, the far-field projection function is applied to the frequency-domain power monitor, which is placed in the substrate 1 µm from the sample surface. Here, only the zeroth diffraction order in the transmission is extracted.

\medskip
\textbf{Supporting Information} \par 
Supporting Information is available through the Wiley Online Library or directly from the authors.

\medskip
\added{\textbf{Data availability}} \par 
\added{The optimization scripts and data are available from the authors upon reasonable request.}

\medskip
\textbf{Acknowledgements} \par 
 The authors thank Prof. Ilya Shadrivov (ANU, Australia) for fruitful discussions.\\
 AR, MJ, EL, BNT, J-SH, and TP acknowledge the support from the Deutsche Forschungsgemeinschaft (DFG, German Research Foundation), Collaborative Research Center (CRC) 1375 NOA "Nonlinear Optics down to Atomic scales" (Project number 398816777), Excellence Cluster (EXC) 2051 "Balance of the Microverse" (Project number 390713860), International Research Training Group 2675 "META-ACTIVE" (project number 437527638), and funding from the Free State of Thuringia and the European Union (project 2025 FGR 0081).\\
 OY, VKM, OMM, and AVK were supported by the Office of Naval Research (ONR) (Grant No. 13001182). The work was performed in part at the Harvard University Center for Nanoscale Systems (CNS); a member of the National Nanotechnology Coordinated Infrastructure Network (NNCI), which is supported by the National Science Foundation under NSF award no. ECCS-2025158.

\medskip
\textbf{Author contributions} \par
\textbf{Anastasia M. Romashkina:} investigation, methodology, visualization, writing – original draft; 
\textbf{Omer Yesilurt:} conceptualization, investigation, methodology, coding;
\textbf{Vahagn K. Mkhitaryan:} methodology, supervision, validation; 
\textbf{Owen M. Matthiessen:} investigation, nanofabrication;
\textbf{Min Jiang:} investigation, writing – review \& editing;
\textbf{Evgeny V. Lyubin:} investigation, writing – review \& editing;
\textbf{Bayarjargal N. Tugchin:} supervision, writing – review \& editing;
\textbf{Isabelle Staude:} funding acquisition, resources;
\textbf{Jer-Shing Huang:} conceptualization, funding acquisition, supervision, writing – review \& editing;
\textbf{Thomas Pertsch:} funding acquisition, supervision, writing – review \& editing;
\textbf{Alexander V. Kildishev:} funding acquisition, methodology, project administration, supervision, writing – review \& editing.

\medskip
\bibliographystyle{MSP}
\bibliography{lib_new}

\newpage

\begin{figure}[h]
\textbf{Table of Contents}\\
\medskip
  \includegraphics{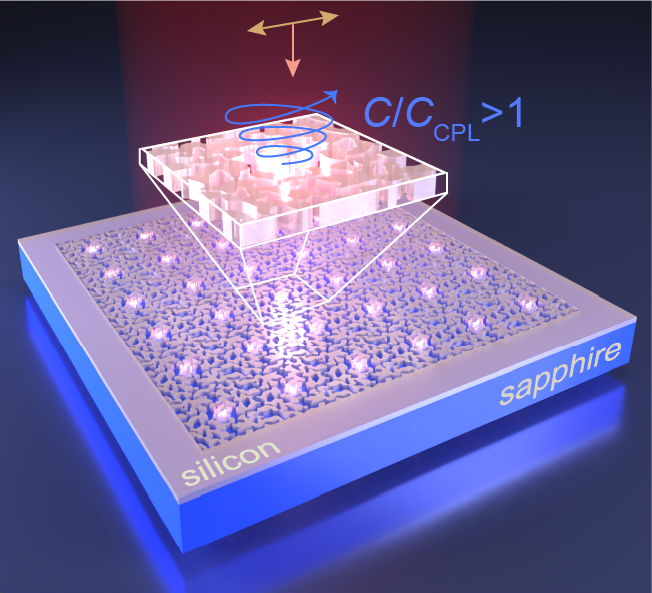}
  \medskip
  \caption*{Linearly polarized incident light with normal incidence creates enhanced optical chirality hotspots on the metasurface. 
  Several orders of magnitude optical chirality enhancement \replaced{offers a promising route toward highly sensitive, compact chiral spectroscopy}{allows for high-sensitivity chiral sensing on the single molecule level}.
  }
\end{figure}

\newpage

\appendix

\sloppy
\pagestyle{fancy}



\title{\textsc{Supporting information}\\ Inverse-Designed Superchiral Hot Spot in Dielectric Meta-Cavity Toward Ultra-Compact Enantioselective Detection}
\maketitle

\author{Anastasia Romashkina$^{\#}$*}
\author{Omer Yesilurt$^{\#}$}
\author{Vahagn Mkhitaryan}
\author{Owen Matthiessen}
\author{Min Jiang}
\author{Evgeny Lyubin}
\author{Bayarjargal N. Tugchin}
\author{Isabelle Staude}
\author{Jer-Shing Huang}
\author{Thomas Pertsch}
\author{Alexander V. Kildishev*}

$^{\#}$Contributed equally

\begin{affiliations}
Dr. O. Yesilurt, Dr. V. K. Mkhitaryan, O. M. Matthiessen, Prof. A. V. Kildishev*\\
Elmore Family School of Electrical and Computer Engineering, Birck Nanotechnology Center, and Purdue Quantum Science and
Engineering Institute, Purdue University, West Lafayette, IN 47907, USA \\ 
Email Address: kildishev@purdue.edu \\

A. M. Romashkina*, Dr. E. V. Lyubin, Dr. B. N. Tugchin, Prof. Isabelle Staude, Prof. T. Pertsch\\
Institute of Applied Physics, Abbe Center of Photonics, Friedrich Schiller University Jena, Albert Einstein Str. 6, 07745 Jena, Germany \\
Email Address: anastasia.romashkina@uni-jena.de
\\

M. Jiang, Prof. J.-S. Huang\\
Research Department of Nanooptics, Leibniz Institute of Photonic Technology, Albert Einstein Str. 9, 07745 Jena, Germany
\\
Prof. I. Staude \\
Institute of Solid State Physics (IFK), Friedrich Schiller University Jena, Helmholtzweg 3, 07743 Jena, Germany

\end{affiliations}


This supplementary document includes sample fabrication details, simulated via the Finite-Difference Time-Domain (FDTD) method transmission spectra of the designed metasurface, additional simulated near-field distributions of the designed and fabricated structures, experimental reflection spectra of the sample, the experimental and numerical results on the characterization of the far-field polarization state of the transmitted light via Stokes parameter formalism for designed and fabricated structures, as well as analysis of sensitivity of design towards small variations of parameters.
\section{FDTD-simulated spectra of the designed structure}

\begin{figure}[ht]
  \centering
  \includegraphics[width=\linewidth]{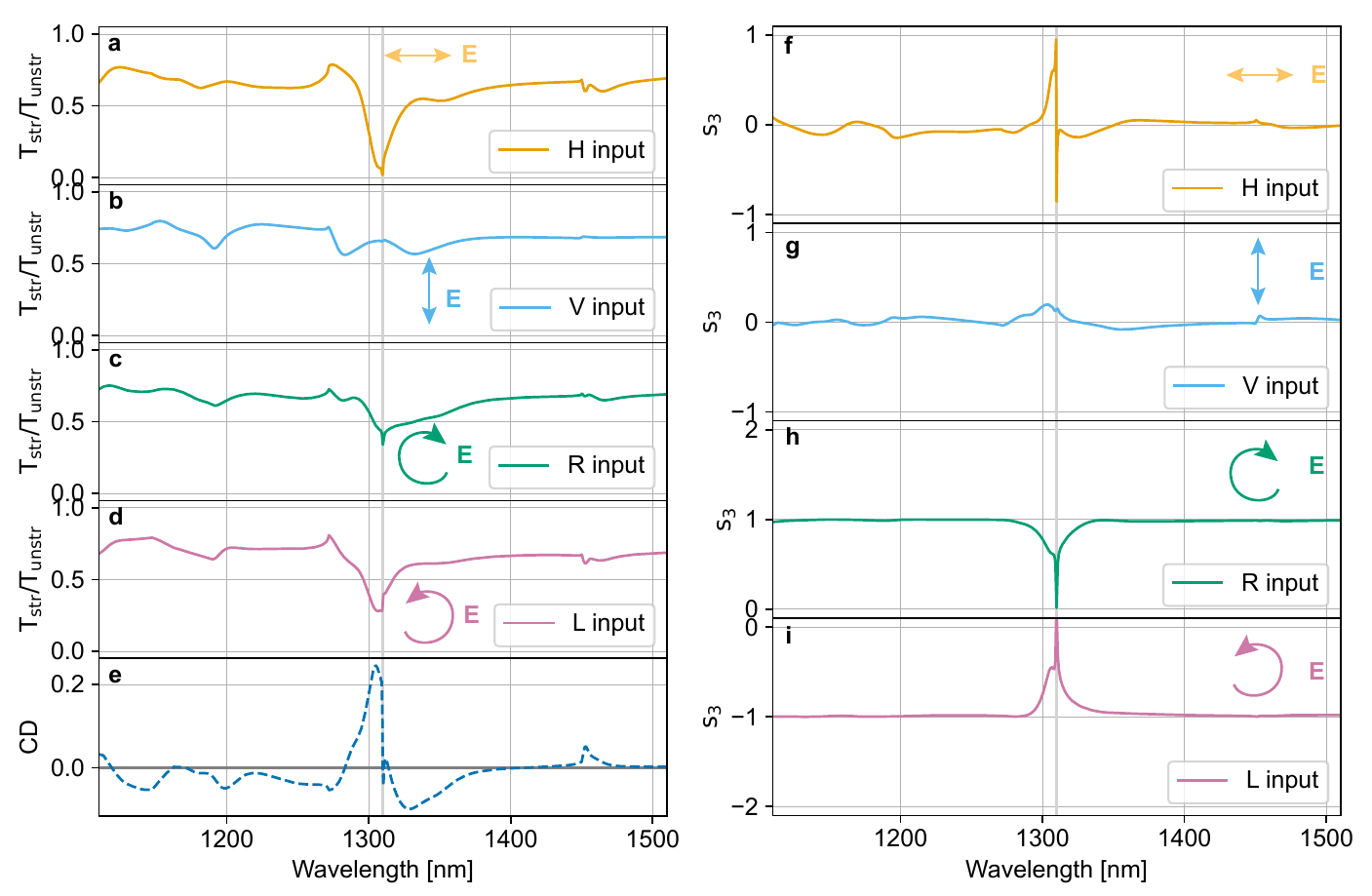}
  \caption{\textbf{FDTD-simulated zeroth diffraction order transmission spectra of the designed structure.} (a-d) Spectra of the total transmission of the metasurface under normal incidence plane wave illumination with H~(a), V~(b), R~(c), and L~(d) polarization. The spectra are normalized to the transmission spectra of an unstructured Si film on the same substrate. (e) The FDTD-simulated CD spectrum in transmission.(f-i) FDTD-simulated (dashed line) normalized third Stokes parameter~$s_3$ spectra of designed metasurface with H-~(f), V-~(g), R-~(h), and L-~(i) input polarizations. The spectra are normalized to the total transmission spectra. Gray vertical line marks 1310~nm wavelength.}
  \label{fig:T_design}
\end{figure}

Figure \ref{fig:T_design}a-d shows FDTD-simulated total transmission spectra of the designed metasurface for horizontal (H), vertical (V), right circular (R), and left circular (L) illumination polarizations. The spectra show prominent sharp features at 1310~nm, which is the target design wavelength of the structure.

Figure \ref{fig:T_design}e shows circular dichroism (CD) spectrum, calculated as $CD = \frac{T_\mathrm{R} - T_\mathrm{L}}{T_\mathrm{R} + T_\mathrm{L}}$, where $T_\mathrm{R}$ and $T_\mathrm{L}$ are transmission coefficients for R- and L- polarized light.

Figure \ref{fig:T_design}f-i shows FDTD-simulated normalized third Stokes parameter $s_3$ spectra for H-~(f), V-~(g), R-~(h), and L-~(i) input polarizations, defined as $s_3 = \frac{I_\mathrm{R}-I_\mathrm{L}}{I_\mathrm{R}+I_\mathrm{L}}$, where $I_\mathrm{R}$ and $I_\mathrm{L}$
are right- and left- circularly polarized components of the transmitted light, respectively.
Third Stokes parameter as a far-field parameter has been shown to be connected to the near-field chirality parameter~\protect \citeSI{poulikakosOptical2016a}.
For the designed structure, a strong resonant enhancement of near-field optical chirality at 1310~nm is observed.
Indeed, an abrupt jump of $s_3$ at the wavelength of 1310~nm under H polarization illumination up to the value of 1 is observed, which means that the light becomes circularly polarized after transmission through the structured film.

\begin{figure}[ht]
  \centering
  \includegraphics[width=0.8\linewidth]{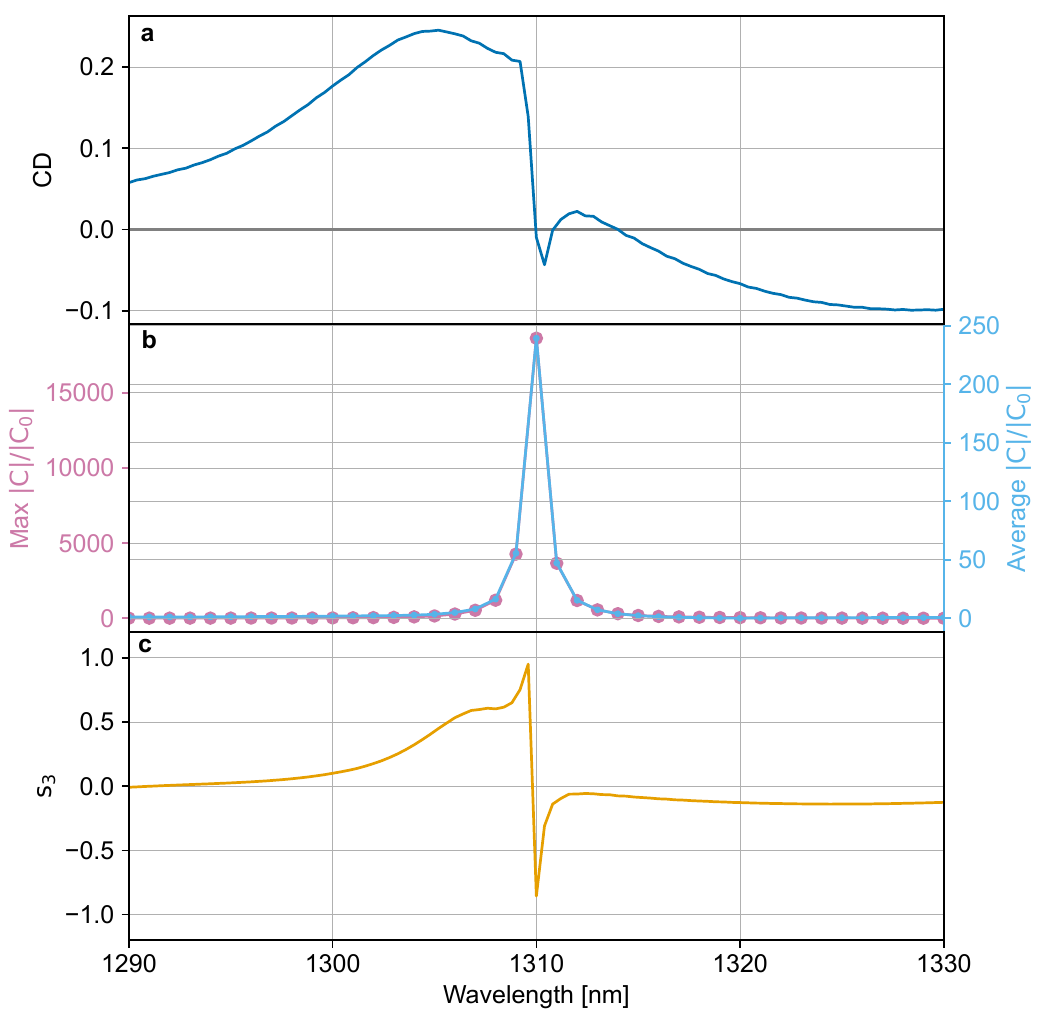}
  \caption{\textbf{FDTD-simulated spectral characterization of the designed resonance} (a) The FDTD-simulated CD spectra in transmission to zeroth diffraction order; (b) Spectra of maximum local chirality value (pink) and average local chirality (blue) across the unit cell on the surface of the structure. The values are normalized to the local chirality value for circularly polarized plane wave in free space; (c) The zeroth order transmission of the H-polarized incident light through the structured Si film: the normalized third Stokes parameter $s_3$ spectrum. The spectrum is normalized to the total transmission spectrum.}
  \label{fig:design_cd_nf_s3}
\end{figure}

Figure \ref{fig:design_cd_nf_s3}a shows the FDTD-simulated CD spectrum for zeroth diffraction order transmission of the designed structure. 
Figure \ref{fig:design_cd_nf_s3}b shows the optical chirality $C$ enhancement on the surface of the structure under H-polarized illumination. At 1310~nm, an abrupt increase in the normalized $C$ is observed. 
Figure \ref{fig:design_cd_nf_s3}c shows the third Stokes parameter in zeroth order transmission. There, a Fano-type resonance is observed at the wavelength of 1310~nm.

\empty
\clearpage

\section{FDTD-simulated near 
field distribution for designed structure: V-pol}

\begin{figure}[ht]
  \centering
  \includegraphics[width=0.8\linewidth]{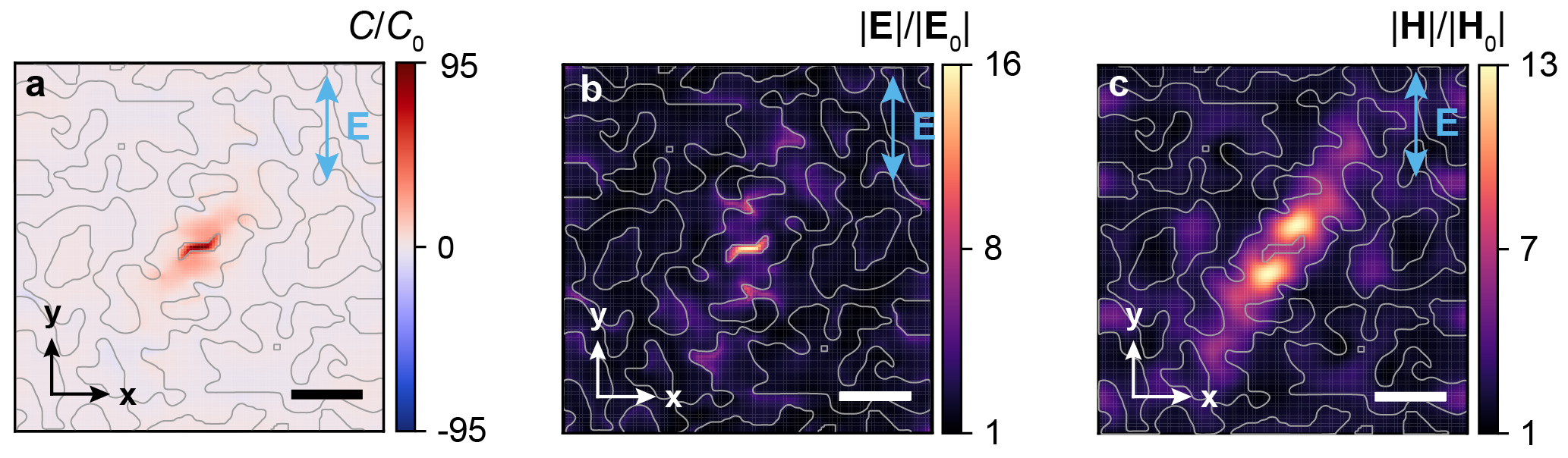}
  \caption{\textbf{FDTD-simulated near fields for designed metasurface model under vertical polarization illumination} (a-c) Simulated pseudo-color maps over the top surface of the unit cell, showing the normalized near-field optical chirality (a), and the normalized magnitudes of the E and H near-fields  (b-c) under monochromatic 1310 nm wavelength plane wave excitation. The values are normalized to the corresponding quantities for circularly polarized light in free space. The gray line outlines the shape of the pattern. Scale bar is 500~nm.}
  \label{fig:nf_v_design}
\end{figure}
\empty
\empty
\clearpage

\section{Fabrication}

\begin{figure}[ht]
  \centering
  \includegraphics[width=\linewidth]{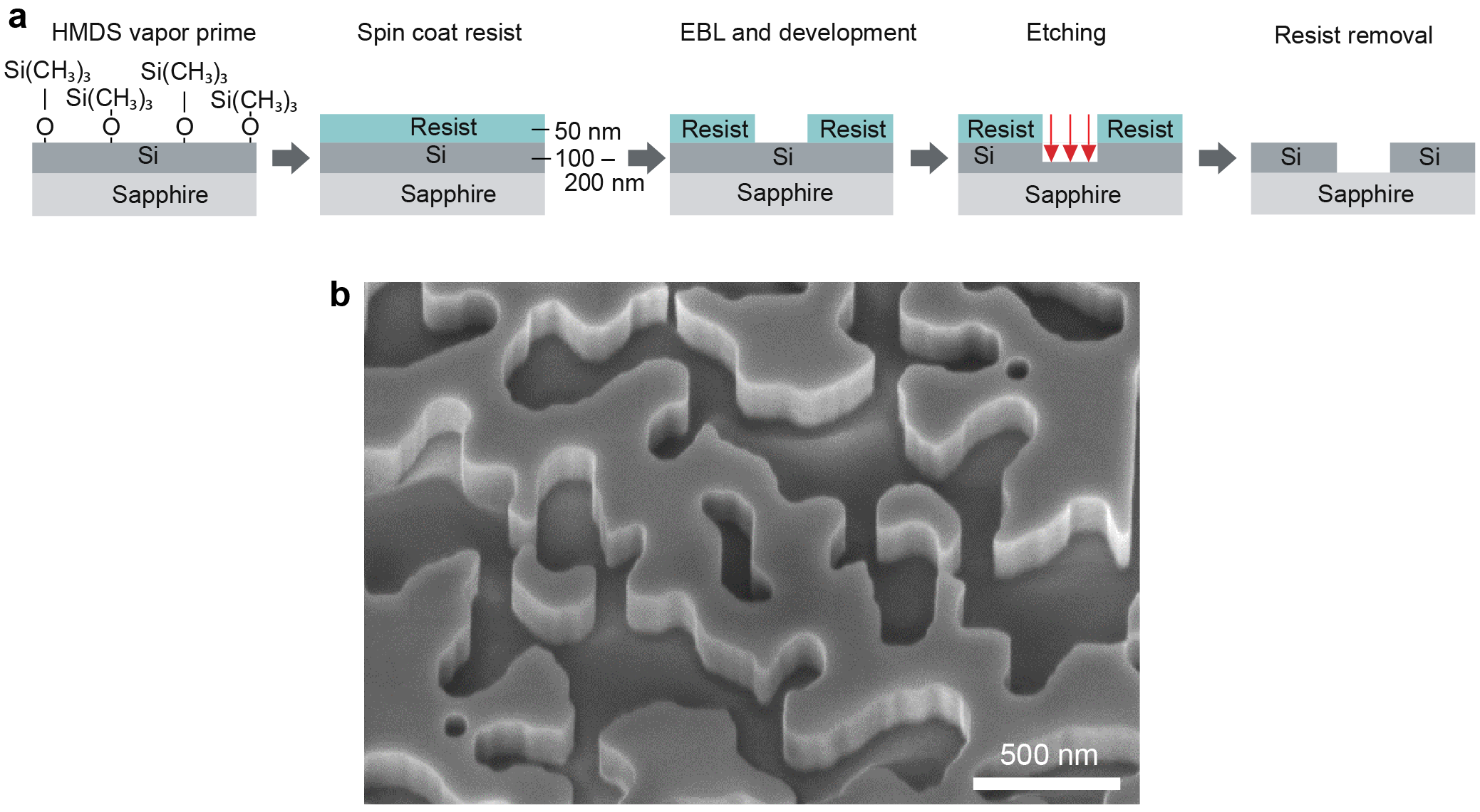}
  \caption{\textbf{Fabrication.} a) Schematic representation of fabrication process steps. Reproduced from~\protect\citeSI{matthiessenOptical2024}. b) SEM image of the fabricated structure with side view. The scale bar is 500 nm. The lateral size of the cavity is approximately 100 nm.}
  \label{fig:sem}
\end{figure}
\empty
\empty
\clearpage

\section{FDTD-simulated near 
field distribution for fabricated structure}

\begin{figure}[ht]
  \centering
  \includegraphics[width=0.8\linewidth]{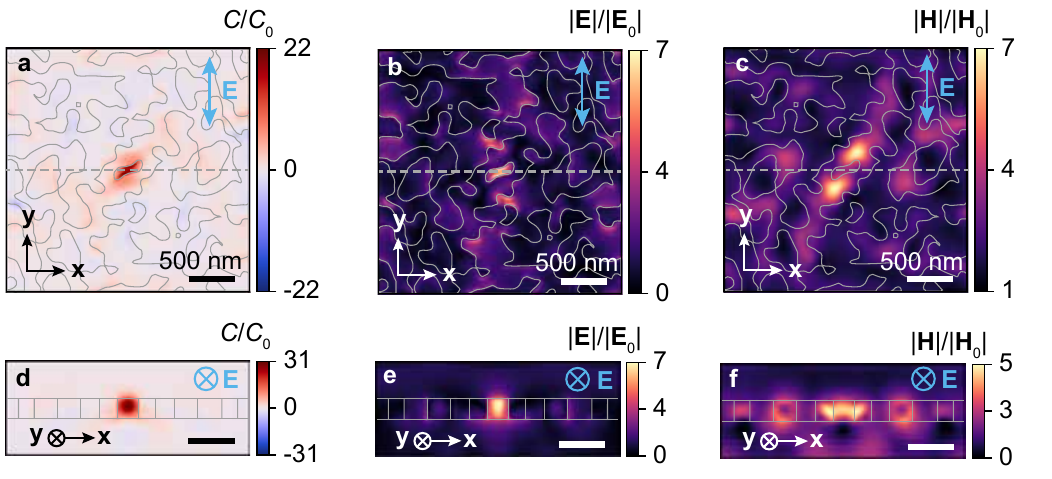}
  \caption{\textbf{FDTD-simulated near fields for fabricated metasurface model under vertical polarization illumination} (a-c) Simulated pseudo-color maps over the top surface of the unit cell, showing the normalized near-field optical chirality (a), and the normalized magnitudes of the E and H near-fields  (b-c). The gray dashed line shows the location of $xz$ plane in (d-f). (d-f) Simulated pseudo-color maps in the $xz$ plane crossing the center of the cavity, showing the normalized near-field optical chirality (d), and the normalized magnitudes of the E and H near-fields (e-f), under monochromatic 1267 nm wavelength plane wave excitation. The values are normalized to the corresponding quantities for circularly polarized light in free space. The gray line outlines the shape of the pattern. Scale bar is 500~nm.}
  \label{fig:nf_v}
\end{figure}

\begin{figure}[h]
  \centering
  \includegraphics[width=0.8\linewidth]{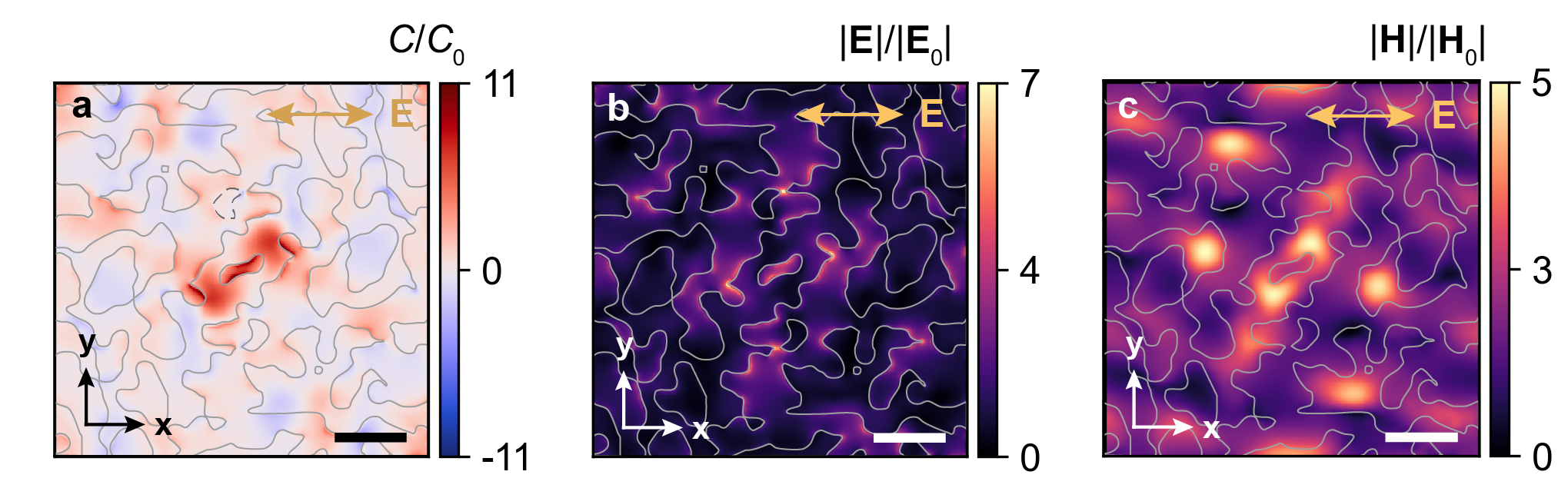}
  \caption{\textbf{FDTD-simulated near fields for a finite, non-periodic model of the fabricated metasurface under H-polarized illumination.} (a-c) Simulated pseudo-color maps over the top surface of the meta-atom, showing the normalized near-field optical chirality (a), and the normalized magnitudes of the E and H near-fields  (b-c). Scale bar is 500~nm.}
  \label{fig:nf_pml}
\end{figure}

\begin{figure}[ht]
  \centering
  \includegraphics[width=0.8\linewidth]{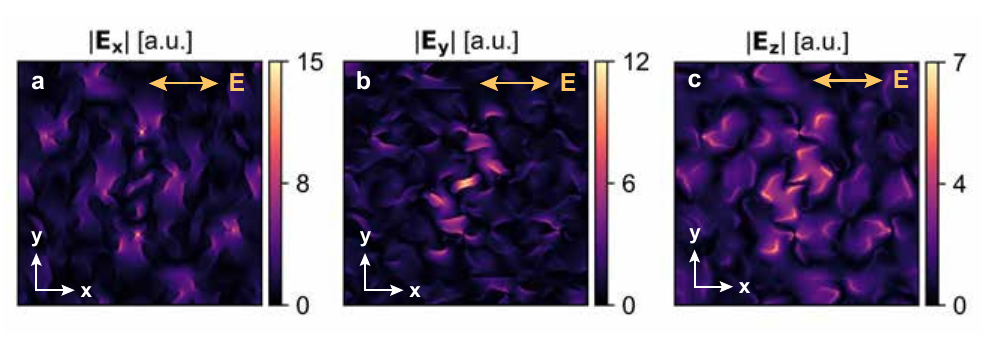}
  \caption{\textbf{FDTD-simulated electric field component maps for fabricated periodic metasurface model under H-polarized illumination.} (a-c) Simulated pseudo-color maps over the top surface of the unit cell, showing the near-field of $x$-component (a), $y$-component (b) and $z$-component (c) of electric field  under monochromatic 1267 nm wavelength plane wave excitation.}
  \label{fig:nf_xyz}
\end{figure}

Figure S4.1 presents the response of the periodic SEM-based unit-cell model under V-polarized excitation and serves as a polarization-control calculation complementary to the H-polarized results shown in Figure 5 of the main text. Figure S4.2 presents the response under H-polarized excitation for a finite, non-periodic SEM-based model, in which the periodic boundary conditions are replaced by perfectly matched layer (PML) boundaries. Comparison with the periodic H-polarized model in Figure 5 provides an estimate of the influence of finite-size and periodicity assumptions on the resonant near-field response.
\empty
\clearpage
\section{Analysis of the transmitted light polarization state}

The Stokes components describe the polarization state of an electromagnetic wave and are defined as

$$
S_0 = I_\mathrm{H} + I_\mathrm{V},~
S_1 = I_\mathrm{H} - I_\mathrm{V},~
S_2 = I_{45} - I_{135},~
S_3 = I_\mathrm{R} - I_\mathrm{L},
$$
where \(I_\mathrm{H}\) and \(I_\mathrm{V}\) are the intensities measured through horizontal and vertical linear polarizers, \(I_{45}\) and \(I_{135}\) are the intensities measured through linear polarizers oriented at \(45^\circ\) and \(135^\circ\), and \(I_\mathrm{R}\) and \(I_\mathrm{L}\) are the intensities of right- and left-circularly polarized components, respectively.  

The Stokes vector
$$
\mathbf{S} = (S_0, S_1, S_2, S_3)^{\mathrm{T}}
$$
fully describes the polarization state of the light.

\begin{figure}[ht]
  \centering
  \includegraphics[width=0.8\linewidth]{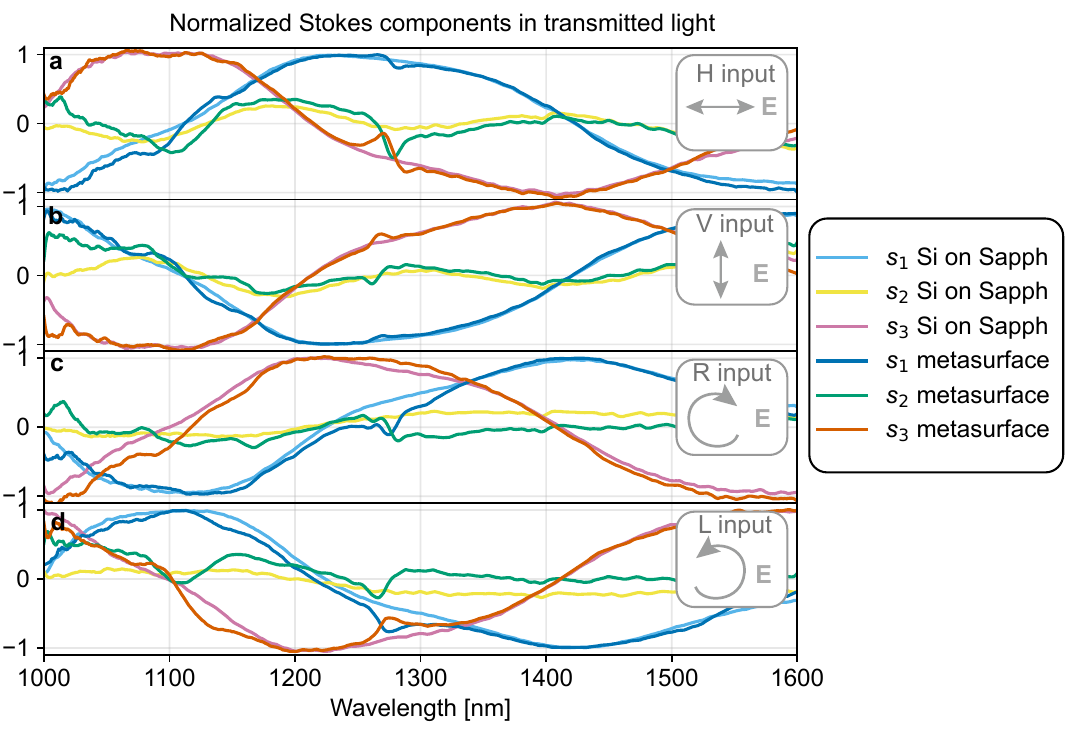}
  \caption{\textbf{Normalized Stokes components for transmitted light.} Stokes components for the transmitted light for H- (a), V- (b), R- (c), and L- (d) input polarizations. Different colors correspond to different components: $s_1$ (light blue), $s_2$ (yellow), $s_3$ (pink) for unstructured Si film on sapphire substrate and $s_1$ (blue), $s_2$ (green), $s_3$ (red) for structured Si film on sapphire substrate (metasurface). }
  \label{fig:s}
\end{figure}

For any input $\mathbf{S}_{\mathrm{in}}$.
\begin{equation}
\mathbf{S}^{\text{out}}_{\text{tot,str}}=\mathbf{M}^{\text{tot}}_{\text{str}}\mathbf{S}_{\mathrm{in}},\qquad
\mathbf{S}^{\text{out}}_{\text{tot,unstr}}=\mathbf{M}^{\text{tot}}_{\text{unstr}}\mathbf{S}_{\mathrm{in}},
\end{equation}
where $\mathbf{M}^{\text{tot}}_{\text{str}}$ and $\mathbf{M}^{\text{tot}}_{\text{unstr}}$ are Mueller matrices of the structured and unstructured Si films on the sapphire substrate, respectively.

Assuming the unstructured film does not affect polarization, so
\begin{equation}
\mathbf{M}^{\text{tot}}_{\text{str}} \propto \mathbf{M}_{\text{str}}\;\mathbf{M}^{\text{tot}}_{\text{unstr}}.
\end{equation}

Solving for the structured film without substrate Mueller matrix:
\begin{equation}
\mathbf{M}_{\text{str}} \approx
\big(\mathbf{M}^{\text{tot}}_{\text{unstr}}\big)^{-1}\,
\mathbf{M}^{\text{tot}}_{\text{str}}\, .\;
\end{equation}

The structure-only output (as if no substrate) is
\begin{equation}
\boxed{\;\mathbf{S}^{\text{out}}_{\text{str}} = \mathbf{M}_{\text{str}}\mathbf{S}_{\mathrm{in}}
\approx \big(\mathbf{M}^{\text{tot}}_{\text{unstr}}\big)^{-1}\,
\mathbf{S}^{\text{out}}_{\text{tot,str}}\, .\;}
\label{eq:stoks}
\end{equation}

\begin{figure}[ht]
  \centering
  \includegraphics[width=0.8\linewidth]{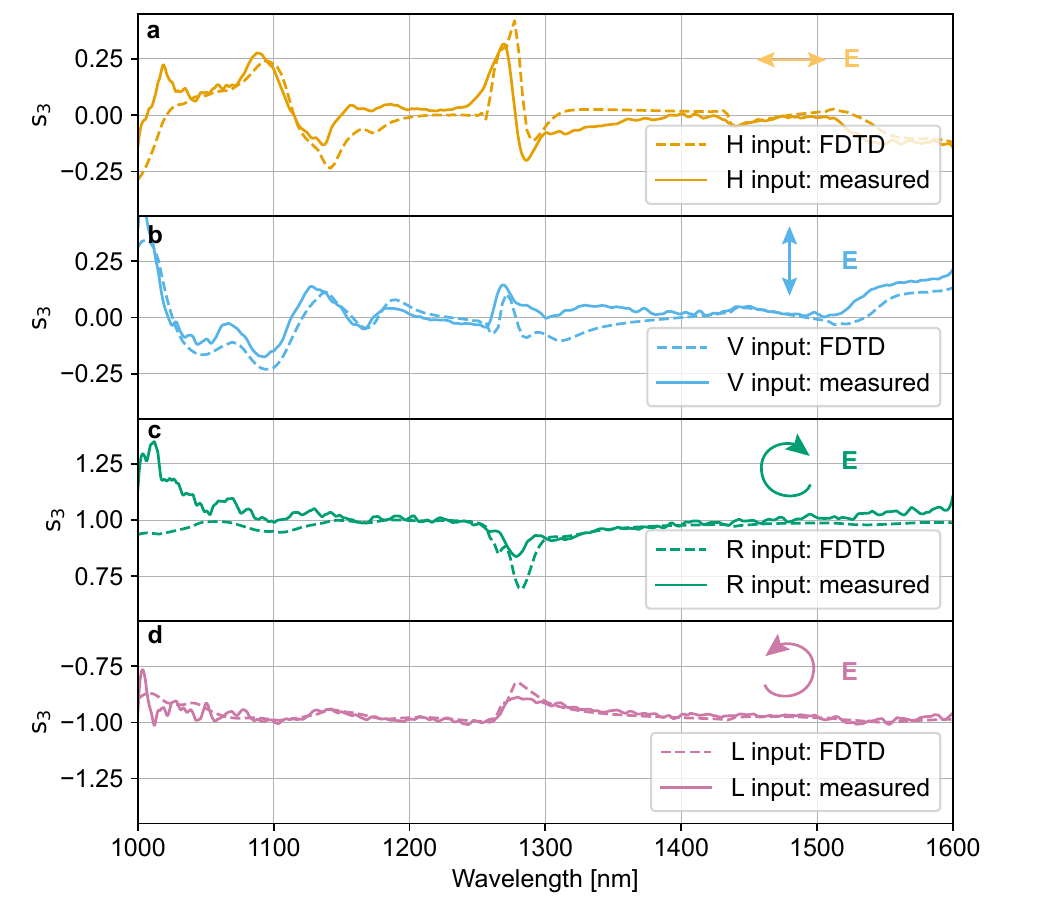}
  \caption{\textbf{Contribution of the structured Si film to the third Stokes parameter in transmission.} The experimental (solid line) and FDTD-simulated (dashed line) normalized third Stokes parameter $s_3$ spectra of structured Si film with H- (a), V- (b), R- (c), and L- (d) input polarizations. The spectra are normalized to the total transmission spectra. For experimental spectra, the structured Si film contribution is calculated with Equation \ref{eq:stoks}.}
  \label{fig:s3}
\end{figure}

\begin{figure}[ht]
  \centering
  \includegraphics[width=0.8\linewidth]{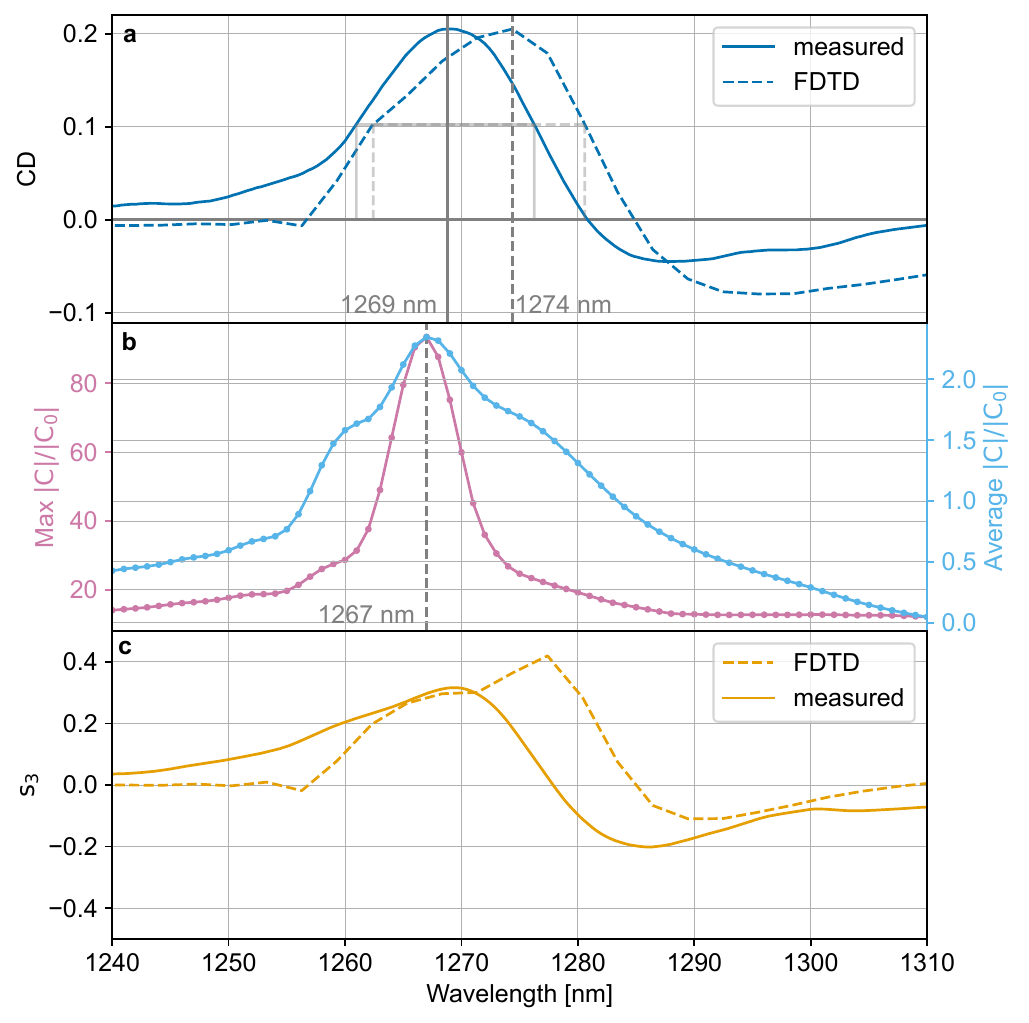}
  \caption{(a) The experimental (solid line) and FDTD-simulated (dashed line) CD spectra in transmission; (b) Spectra of maximum local chirality value (pink dashed line) and average local chirality across the unit cell on the surface of the structure. The values are normalized to the local chirality value for circularly polarized plane wave in free space; (c) The experimental (solid line) and FDTD-simulated (dashed line) normalized third Stokes parameter $s_3$ spectra of structured Si film with H- input polarization. The spectra are normalized to the total transmission spectra. For experimental spectra, the structured Si film contribution is calculated with Equation \ref{eq:stoks}.}
  \label{fig:cd_nf_s3}
\end{figure}

\empty
\clearpage
\empty
\empty
\clearpage
\empty
\newpage

\section{Reflection spectra of the fabricated structure}

The reflection spectra were measured experimentally using a microscope-based spectroscopy setup with the 10$\times$ objective for illumination and collection of the light. 

To simulate the non-zero numerical aperture (NA) by FDTD calculations, a stack of broadband simulations with oblique incidence angles was performed. The reflection spectra were then combined as a weighted sum \citeSI{novotnyPrinciples2012}, including the correction of wavelength that needs to be applied when using Bloch boundary conditions and broadband simulation \citeSI{Lumerical_si}.

\begin{figure}[ht]
  \centering
  \includegraphics[width=0.8\linewidth]{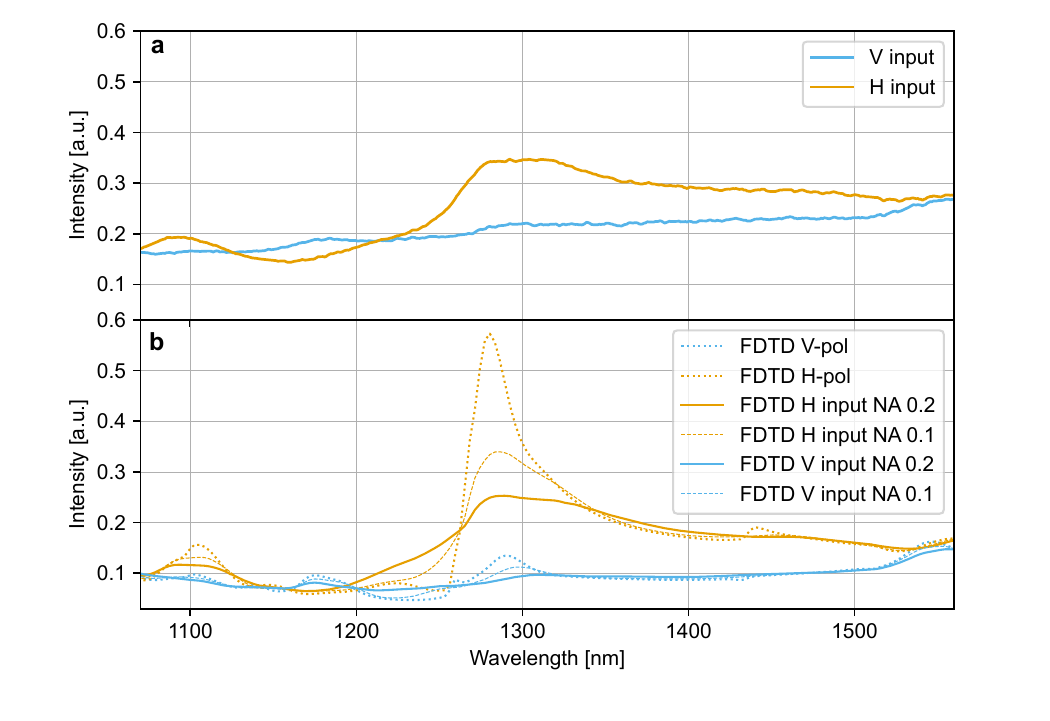}
  \caption{\textbf{Reflection spectra of the sample} a) Experimental reflection spectra of the metasurface for H- (red line) and V- (blue line) polarizations. b) Numerically modelled via FDTD method reflection spectra of the sample with plane wave illumination (dotted lines), 0.1 NA illumination (dashed lines), 0.2 NA illumination (solid lines) for H- and V- polarizations, represented as red and blue lines, respectively.}
  \label{fig:refl}
\end{figure}
\empty
\empty
\clearpage
\section{Robustness and sensitivity analysis of the design}

\noindent\textbf{(a)}

\begin{tabular}{ |p{5cm}|p{3cm}|p{3cm}|p{3cm}|p{3cm}|}
\hline
& Monochromatic design model & Design model & Design model with Si film thickness of +5\% & Design model with period 2605 nm  \\
\hline
Simulation bandwidth [nm] & 1 & 100 & 100 & 100\\
\hline
Pattern model & design & design & design & design\\
\hline
Si thickness [nm] & 200 & 200 & 210 & 200\\
\hline
Period [nm]& 2600 & 2600 & 2600 & 2605\\
\hline
Si refr. index & $3.5 + i\cdot10^{-5}$ & $3.5 + i\cdot10^{-5}$ & $3.5 + i\cdot10^{-5}$& $3.5 + i\cdot10^{-5}$\\
\hline
Sapph refr. index & 1.762555 & 1.762555 & 1.762555 & 1.762555 \\
\hline
Resonance wavelength [nm] & 1310 & 1310 & 1319 & 1304 \\
\hline
Max $|C|/|C_0|$ & 18635 & 9542 & 3349 & 4357 \\
\hline
Max $|\mathbf{E}|/|\mathbf{E}_0|$ & 211 & 125 & 75 & 96 \\
\hline
Max $|\mathbf{H}|/|\mathbf{H}_0|$ & 169 & 133 & 79 & 87 \\
\hline
\end{tabular}

\vspace{1em}
\noindent\textbf{(b)}

\begin{tabular}{ |p{5cm}|p{3cm}|p{3cm}|p{3cm}|p{3cm}|}

\hline
& Design model & Design pattern imported as an image & Design pattern imported as image with different material data & Perturbed design pattern\\
\hline
Simulation bandwidth [nm] & 100 & 100 & 100 & 100 \\
\hline
Pattern model & design & design image & design image & design image with 5\% change \\
\hline
Si thickness & 200 & 200 & 200 & 200 \\
\hline
Period & 2600 & 2600 & 2600 & 2600 \\
\hline
Si refr. index  &$3.5 + i\cdot10^{-5}$ & $3.5 + i\cdot10^{-5}$ & 3.502 at 1310~nm (dispersive)& $3.5 + i\cdot10^{-5}$\\
\hline
Sapph refr. index & 1.762555 & 1.762555 & 1.743 at 1310~nm (dispersive) & 1.762555 \\
\hline
Resonance wavelength [nm] & 1310 & 1301 & 1295 & 1297 \\
\hline
Max $|C|/|C_0|$  & 9542 & 2818 & 2766 & 933 \\
\hline
Max $|\mathbf{E}|/|\mathbf{E}_0|$ & 125 & 73 & 67 & 40 \\
\hline
Max $|\mathbf{H}|/|\mathbf{H}_0|$ & 133 & 69 & 70 & 42 \\
\hline

\end{tabular}

{\captionof{table}{
\textbf{Sensitivity of the simulated optical response to numerical,
geometrical, and material variations.}
\textbf{(a)} Comparison of monochromatic and finite-bandwidth simulations,
including the effects of a 5\% increase in the Si-film thickness and a
5~nm increase in the structure period.
\textbf{(b)} Comparison of the original design model with image-imported
patterns, including dispersive material properties and a 5\% geometrical
perturbation. For each configuration, the resonance wavelength and the
maximum normalized optical chirality, electric-field amplitude, and
magnetic-field amplitude are reported.
}}

\vspace{1 cm}

The results summarized in Table~S1 assess the sensitivity of the designed structure to the simulation bandwidth, geometric parameters, pattern representation, and material properties. The monochromatic model produces the largest field enhancements on the surface of the structure, with maximum values of $|C|/|C_0|=18,635$, $|\mathbf{E}|/|\mathbf{E}_0|=211$, and $|\mathbf{H}|/|\mathbf{H}_0|=169$ at a resonance wavelength of 1310~nm. When a 100~nm simulation bandwidth is considered, the resonance wavelength remains at 1310~nm, whereas the corresponding maximum enhancements decrease to 9542, 125, and 133, respectively. This difference indicates that the exceptionally high monochromatic response is reduced when the device performance is evaluated over a finite spectral range.

The design is also sensitive to small variations in its geometric parameters. Increasing the Si-film thickness by 5\%, from 200 to 210~nm, shifts the resonance from 1310 to 1319~nm and decreases the maximum $|C|/|C_0|$ from 9542 to 3349. The electric- and magnetic-field enhancements decrease from 125 and 133 to 75 and 79, respectively. Similarly, increasing the period from 2600 to 2605~nm shifts the resonance to 1304~nm and reduces the maximum $|C|/|C_0|$ to 4357, while the electric- and magnetic-field enhancements decrease to 96 and 87. Thus, even relatively small deviations in the film thickness or lattice period can noticeably modify both the resonance position and its enhancement strength.

Additional differences arise when the optimized pattern is imported as an image rather than represented directly by the design model. Image-based pattern import shifts the resonance from 1310 to 1301~nm and reduces the maximum $|C|/|C_0|$ from 9542 to 2818. The corresponding electric- and magnetic-field enhancements decrease to 73 and 69. Including dispersive material data results in a further resonance shift to 1295~nm but the field enhancement remains almost unchanged.

Further degradation is observed for the design topology perturbed by 5\% point-by-point towards the geometry, extracted from SEM  image. In this case, the resonance occurs at 1297~nm, while the maximum $|C|/|C_0|$ decreases to 933 and the electric- and magnetic-field enhancements decrease to 40 and 42. 
Overall, these results show that the maximum enhancement is sensitive to geometric perturbations and meshing artifacts during pattern construction, as well as refractive index variations. Accurate fabrication and correct reproduction of the optimized geometry are therefore important for preserving the predicted enhancement magnitude.
\empty
\empty
\clearpage
\section{The Neural network architecture for the first step of the inverse design algorithm}

\begin{figure}[ht]
  \centering
  \includegraphics[width=\linewidth]{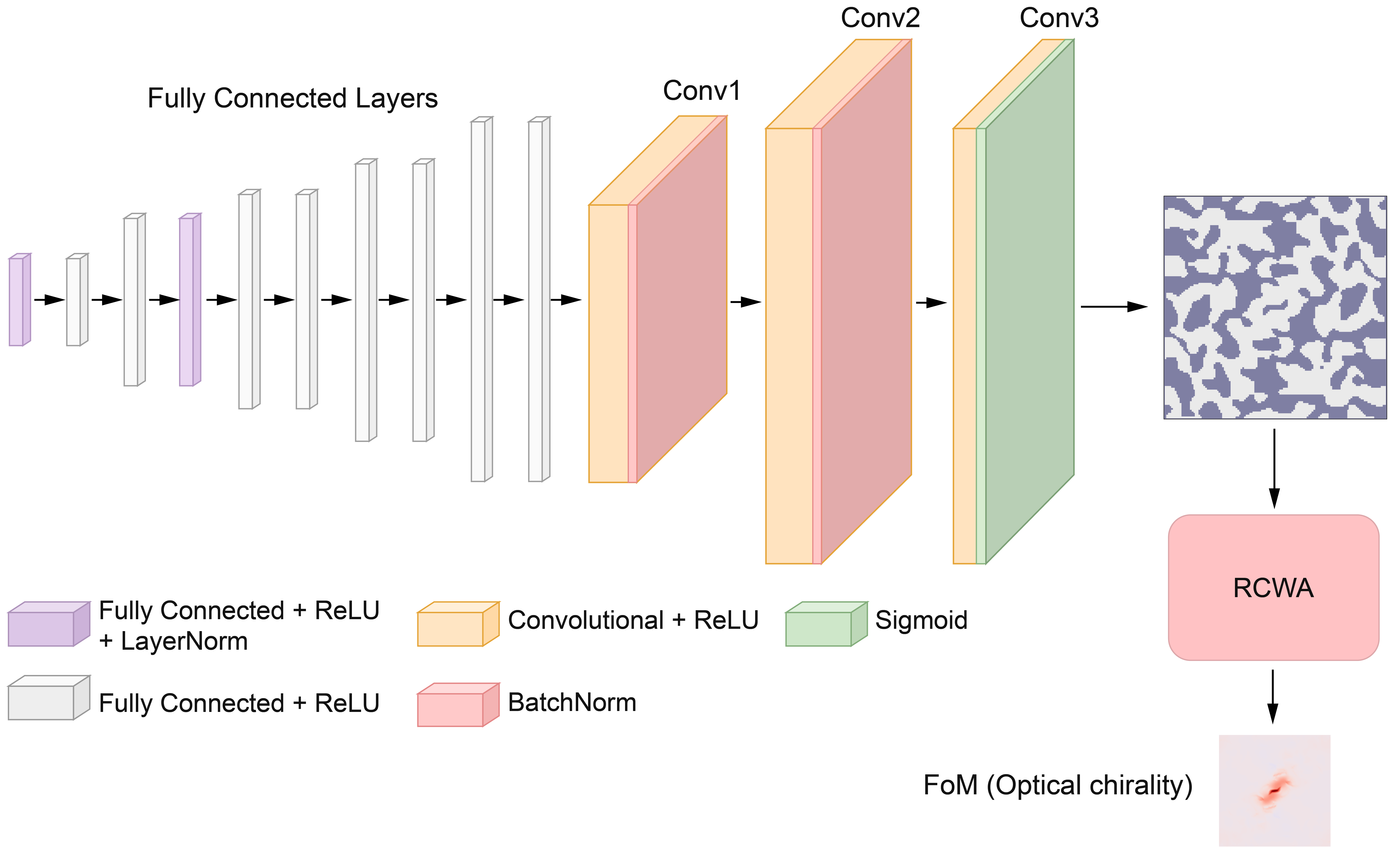}
  \caption{\textbf{The NN-driven inverse design scheme for superchiral metasurface unit cells.} The NN progressively expands a noise vector into a complex topology of the correct dimensions. The EM response of the proposed design is then calculated using the RCWA method. After the loss is obtained, the gradients are backpropagated through the fully differentiable RCWA into the NN for parameter updating. This process iterates until convergence is reached in the figure of merit or based on user input. Adapted from~\protect\citeSI{yesilurtPhysics2024}.}
  \label{fig:NN_si}
\end{figure}

\bibliographystyleSI{MSP}
\bibliographySI{SI}

\end{document}